\documentclass[iop,apj]{emulateapj}
\usepackage[figuresright]{rotating}
\begin{document}

\shorttitle{Twenty New TeV Blazars}

\title{First-Epoch VLBA Imaging of Twenty New TeV Blazars}

\author{B.~Glenn~Piner\altaffilmark{1,2} \& Philip~G.~Edwards\altaffilmark{3}}
\altaffiltext{1}{Department of Physics and Astronomy, Whittier College,
13406 E. Philadelphia Street, Whittier, CA 90608, USA; gpiner@whittier.edu}
\altaffiltext{2}{Jet Propulsion Laboratory, California Institute of Technology,
4800 Oak Grove Drive, Pasadena, CA 91106, USA}
\altaffiltext{3}{CSIRO Astronomy and Space Science, Australia Telescope National Facility,
PO Box 76, Epping, NSW 1710, Australia}

\begin{abstract}
We present Very Long Baseline Array (VLBA) images of 
20 TeV blazars (HBLs) not previously well-studied on the parsec scale.
Observations were made between August and December 2013, at a frequency of 8.4~GHz.
These observations represent the first epoch of a VLBA monitoring campaign on these blazars,
and they significantly increase the fraction of TeV HBLs studied with high-resolution imaging.
The peak VLBI flux densities of these sources range from $\sim10$ to $\sim100$~mJy~bm$^{-1}$,
and parsec-scale jet structure is detected in all sources. 
About half of the VLBI cores are resolved, with brightness temperature
upper limits of a few times $10^{10}$~K, and we find that a 
brightness temperature of $\sim2\times10^{10}$~K
is consistent with the VLBI data for all but one of the sources.
Such brightness temperatures do not require any relativistic beaming to
reduce the observed value below commonly invoked intrinsic limits; however, the lack of
detection of counter-jets does place a modest limit on the bulk Lorentz factor of $\gamma\gtrsim2$.
These data are thus consistent with a picture where
weak-jet sources like the TeV HBLs develop significant velocity structures on parsec-scales.
We also extend consideration to the full sample of TeV HBLs
by combining the new VLBI data with VLBI and gamma-ray data from the literature.
By comparing measured VLBI and TeV fluxes to samples with intrinsically uncorrelated luminosities generated by
Monte Carlo simulations, we find a marginally significant correlation 
between the VLBI and TeV fluxes for the full TeV HBL sample.
\end{abstract}


\section{Introduction}
\label{intro}
The number of TeV blazars has grown rapidly in recent years
(see, e.g., the reviews of Holder 2012, 2014)
with 54 TeV blazars currently known\footnote{\url{http://tevcat.uchicago.edu/}}.
The vast majority of these (44 of 54, or about 80\%)
belong to the blazar sub-class of high-frequency peaked BL Lac objects, or
HBLs. Several of these TeV HBLs have displayed remarkable variability in
their TeV gamma-ray emission on time scales as short as a few minutes
(e.g., Aharonian et al.\ 2007; Albert et al.\ 2007; Sakamoto et al.\ 2008).
Although various explanations have been proposed for such rapid variability
(e.g., Begelman et al.\ 2008; Nalewajko et al.\ 2011; Narayan \& Piran 2012; Barkov et al.\ 2012),
they share the common feature of high bulk Lorentz factors of at least $\gtrsim$25
for the gamma-ray emitting plasma in their relativistic jets. 
High bulk Lorentz factors and Doppler factors are
also required to model TeV blazar spectral energy distributions
(e.g., Tavecchio et al.\ 2010), particularly in the case of one-zone models.
For example, fitting the SED of the TeV blazar PKS~1424+240 with a one-zone
model yields a Doppler factor of $\delta\sim100$ (Aleksi{\'c}, et al.\ 2014a).

Direct imaging of the jets of these blazars on parsec-scales
requires VLBI. Most HBLs are relatively faint in the radio, so the TeV HBLs
are not well represented in large VLBI monitoring projects,
such as MOJAVE (Lister et al.\ 2009). We have
previously reported multi-epoch VLBI kinematic results for 11
established TeV HBLs (Piner et al.\ 2010; Tiet et al.\ 2012).
A major result of those kinematic analyses was the absence of any
rapidly moving features in the jets of those blazars; all components in all
11 sources were either stationary or slowly moving ($\lesssim$1$c$). Slow
apparent speeds of VLBI components in specific TeV HBLs has been
confirmed by numerous other studies (e.g., Giroletti et al.\ 2004a; Lico et al.\ 2012; Blasi et al.\ 2013;
Aleksi{\'c} et al.\ 2013; Richards et al.\ 2013), although note that TeV-detected
intermediate-peaked BL Lac objects (IBLs), such as 3C 66A and BL Lac,
do show apparently superluminal components (e.g., Britzen et al.\ 2008).
While effects other than slow bulk motion can produce
slow apparent speeds of components, the complete absence of any
rapidly moving features in all of these jets, after as much as 20 years of
VLBA monitoring (for Mrk 421 and Mrk 501), and even after powerful
flares (Richards et al.\ 2013), is quite distinct from the behavior of other types of
gamma-ray blazars, which show frequent superluminal ejections
(e.g., Lister et al.\ 2009; Marscher 2013). 
Taken together with other measured radio properties, such as
the brightness temperatures and core dominance
(Giroletti et al.\ 2004b; Lister et al.\ 2011), the VLBI data
imply only modest bulk Lorentz factors and Doppler factors in the parsec-scale radio jets
of these TeV HBLs. (Note that, because the sources appear one-sided on parsec
scales, the VLBI data do require that the sources be at least moderately
relativistic.) This discrepancy between the Doppler and Lorentz factors
estimated from the gamma-ray data and the radio date has been
referred to as the ``Doppler Crisis'' of TeV blazars.

A natural explanation for the Doppler Crisis is that the radio and
gamma-ray emission may be produced in different parts of the jet with
different bulk Lorentz factors. Several variations of such a multi-component jet have
been proposed including decelerating jets (Georganopoulos \& Kazanas 2003),
spine-sheath structures (Ghisellini et al.\ 2005), 
`minijets' within the main jet (Giannios et al.\ 2009), and 
faster moving leading edges of blobs (Lyutikov \& Lister 2010), 
but they all require that the jets of HBLs contain significant velocity structures. 
Some of these velocity structures, such as a fast spine and slower layer,
may under certain conditions produce observable signatures in VLBI
images, such as limb brightening of the transverse jet structure. Limb
brightening has indeed been observed in VLBI images of the bright TeV blazars Mrk 421
and Mrk 501 (e.g., Giroletti et al.\ 2004a, 2006, 2008; Piner et al.\ 2009, 2010;
Croke et al.\ 2010; Blasi et al.\ 2013).

These arguments for velocity structures in the jets of TeV HBLs are
independently supported by developments in radio-loud AGN unification 
(Meyer et al.\ 2011, 2013a; see also Ghisellini et al.\ 2009).
In that unification work, radio-loud AGN are divided into two
distinct sub-populations that constitute a `broken power
sequence'. The `weak' jet sub-population resulting from  
inefficient accretion modes (and corresponding to HBLs when
viewed at a small angle) follows a de-beaming curve that requires
velocity gradients in the jets, such as a decelerating or
spine-sheath jet; see also the similar arguments in earlier unification work
(Chiaberge et al.\ 2000). The TeV HBLs may thus represent the small
viewing angle peak of a distinct radio-loud population with 
both fundamentally different jet structure and accretion mode from the more powerful blazars.
If this is the case, then obtaining more information on the parsec-scale
structure of these sources through high-resolution imaging is quite important.

We are presently taking advantage of both the rapidly growing TeV
blazar source list and the recently upgraded sensitivity of the Very Long
Baseline Array (VLBA) to significantly expand our previous work on the parsec-scale
structure of TeV HBLs (e.g., Piner et al.\ 2010; Tiet et al.\ 2012).
Here, we present first-epoch VLBA images of twenty newer TeV HBLs
discovered during the years 2006 to 2013, several of which had never been imaged with VLBI.
This represents the first stage of a multi-epoch VLBA monitoring program
on these sources designed to provide parsec-scale kinematic and structural information
on nearly the complete sample of TeV HBLs.
In $\S$~\ref{observations} we describe the source selection and observations,
in $\S$~\ref{results20} we present the results for these observations, and in $\S$~\ref{resultsall}
we extend consideration to the full set of TeV HBLs.
Final discussion and conclusions are given in $\S$~\ref{discussion}.

\section{Observations}
\label{observations}
\subsection{Source Selection}
\label{sourceselection}
We have been conducting VLBA observations of TeV-detected HBLs since the 
discovery of the first two TeV blazars (Mrk~421 and Mrk~501) in the 1990s, in order to study their jet physics
through high-resolution parsec-scale imaging (see $\S$~\ref{intro}). 
Our complete candidate source list is thus the 44 HBLs listed as detections
in the TeVCat catalog\footnote{\url{http://tevcat.uchicago.edu/}} as of this writing.
From those 44 sources we excluded the following for the observations in this paper:
\begin{enumerate}
\item{Eleven sources reported as TeV detections before 2007
for which we have already published multi-epoch VLBA observations: six 
of these sources are discussed by Piner et al.\ (2010),
and an additional five by Tiet et al.\ (2012).}
\vspace*{-0.05in}
\item{Seven sources with sufficient multi-epoch VLBA data in the MOJAVE monitoring program
\footnote{\url{http://www.physics.purdue.edu/astro/MOJAVE/\\allsources.html}}.}
\vspace{-0.05in}
\item{Three sources which are below $-40\arcdeg$ declination, and thus difficult to image with the VLBA.}
\vspace{-0.05in}
\item{Two sources which were detected too recently (after 2013) to be included in this work.}
\vspace{-0.05in}
\item{The low brightness temperature source HESS~J1943+213 (Gab{\'a}nyi et al.\ 2013).}
\end{enumerate} 
These exclusions are shown in tabular from in Table~\ref{selecttab}, 
and they leave 20 HBLs (or nearly half of the full sample)
that were all reported as new detections by the TeV telescopes between 2006 and 2013,
and that have not yet been studied with multi-epoch VLBI imaging by any program.
The goal of the observations presented here is to provide high-dynamic range single-epoch images
of these 20 sources, as a precursor to a multi-epoch monitoring program to study the jet kinematics.

\begin{table*}[!t]
\begin{center}
{\normalsize \caption{Sample Selection}
\label{selecttab}
\begin{tabular}{l c c l c c} \tableline \tableline \\[-5pt]
\multicolumn{1}{c}{Source$^{a}$} & Included$^{b}$ & Reason$^{c}$ & 
\multicolumn{1}{c}{Source$^{a}$} & Included$^{b}$ & Reason$^{c}$  \\ \tableline \\[-5pt]
SHBL~J001355.9$-$185406 & Y & ... & Markarian~421   & N & 1   \\ [2pt]
KUV~00311$-$1938        & Y & ... & Markarian~180   & N & 1   \\ [2pt]
1ES~0033+595            & Y & ... & RX~J1136.5+6737 & N & 4   \\ [2pt]
RGB~J0136+391           & Y & ... & 1ES~1215+303    & N & 2   \\ [2pt]
RGB~J0152+017           & Y & ... & 1ES~1218+304    & N & 1   \\ [2pt]
1ES~0229+200            & Y & ... & MS~1221.8+2452  & Y & ... \\ [2pt]
PKS~0301$-$243          & N & 2   & 1ES~1312$-$423  & N & 3   \\ [2pt]
IC~310                  & N & 2   & PKS~1424+240    & N & 2   \\ [2pt]
RBS~0413                & Y & ... & H~1426+428      & N & 1   \\ [2pt]
1ES~0347$-$121          & Y & ... & 1ES~1440+122    & Y & ... \\ [2pt]
1ES~0414+009            & Y & ... & PG~1553+113     & N & 1   \\ [2pt]
PKS~0447$-$439          & N & 3   & Markarian~501   & N & 1   \\ [2pt]
1ES~0502+675            & Y & ... & H~1722+119      & Y & ... \\ [2pt]
PKS~0548$-$322          & Y & ... & 1ES~1727+502    & N & 2   \\ [2pt]
RX~J0648.7+1516         & Y & ... & 1ES~1741+196    & Y & ... \\ [2pt]
1ES~0647+250            & Y & ... & HESS~J1943+213  & N & 5   \\ [2pt]
RGB~J0710+591           & Y & ... & 1ES~1959+650    & N & 1   \\ [2pt]
1ES~0806+524            & N & 2   & PKS~2005$-$489  & N & 3   \\ [2pt]
RBS~0723                & N & 4   & PKS~2155$-$304  & N & 1   \\ [2pt]
1RXS~J101015.9$-$311909 & Y & ... & B3~2247+381     & Y & ... \\ [2pt]
1ES~1011+496            & N & 2   & 1ES~2344+514    & N & 1   \\ [2pt]
1ES~1101$-$232          & N & 1   & H~2356$-$309    & N & 1   \\ \tableline \\[-10pt]
\end{tabular}}
\end{center}
{\bf Notes.}\\
$a$: Source names are the so-called `Canonical Name' used by TeVCat.\\
$b$: Whether or not the source is included in the new observations for this paper.\\
$c$: Reason for exclusion: 1:~Monitored in our previous work, 2:~in MOJAVE program,
3:~too far south, 4:~detection too recent, 5:~low brightness temperature (see $\S$~\ref{sandtb}).\\
\end{table*}

Single-epoch pilot images of 8 of these 20 sources, obtained prior to the VLBA sensitivity upgrade (Romney et al.\ 2009),
were presented by Piner \& Edwards (2013). For those 8 sources, we present
in this paper the first epoch of a multi-epoch monitoring series obtained after the sensitivity upgrade,
and with images of significantly higher dynamic range compared to those in Piner \& Edwards (2013). 
For the remaining 12 sources, we present
the pilot images made to assess suitability for multi-epoch monitoring, all of which were made
subsequent to the VLBA sensitivity upgrade.

The VLBI and gamma-ray properties of the entire sample of 44 TeV HBLs are discussed later in this paper
(see $\S$~\ref{resultsall} and Table~\ref{alltab}).

\subsection{Details of Observations}
\label{obsdetails}
Details of the observing sessions are given in Table~\ref{obstab}.
All observations were made at an observing frequency of 8.4~GHz (4~cm), because this provides the
optimum combination of angular resolution and sensitivity for these fainter sources.
All observations used the full 2 Gbps recording rate of the VLBA, and were
made using the polyphase filterbank (PFB) observing system of the Roach Digital Backend (RDBE), in its
dual-polarization configuration of eight contiguous 32~MHz channels at matching frequencies in each polarization.
Although dual-polarization was recorded, only total intensity (Stokes I) was calibrated and imaged,
because of the likely sub-millijansky level of polarized flux from most of these sources.

We used phase-referencing for three of the fainter targets:
SHBL~J001355.9$-$185406, 1ES~0347$-$121, and 1RXS~J101015.9$-$311909; both because
their correlated flux densities were uncertain, and to obtain precise milliarcsecond-scale
positions because they were not in the VLBA input catalog.
These data were phase-referenced to the ICRF sources J0015$-$1812, J0351$-$1153, and J1011$-$2847
respectively, all of which had separations of less than $3\arcdeg$ from the target sources.
Based on the ICRF positions of the calibrator sources, we derived (J2000) positions of
R.A.~=~00$^{\rm {h}}$13$^{\rm {m}}$56.043$^{\rm {s}}$, decl.~=~$-$18$\arcdeg$54$'$06.696$''$ for SHBL~J001355.9$-$185406,
R.A.~=~03$^{\rm {h}}$49$^{\rm {m}}$23.186$^{\rm {s}}$, decl.~=~$-$11$\arcdeg$59$'$27.361$''$ for 1ES~0347$-$121, and
R.A.~=~10$^{\rm {h}}$10$^{\rm {m}}$15.979$^{\rm {s}}$, decl.~=~$-$31$\arcdeg$19$'$08.408$''$ for SHBL~J001355.9$-$185406,
which we expect to be accurate to a few milliarcseconds.
Although those observations were done in phase-referencing mode, all three sources
were bright enough for fringe-fitting, and the fringe-fit data were used 
in the subsequent imaging.

\begin{table*}[!t]
\begin{center}
{\normalsize \caption{Observation Log}
\label{obstab}
\begin{tabular}{l c c c c} \tableline \tableline \\[-5pt]
\multicolumn{1}{c}{Date} & Observation & Observing & Excluded & \multicolumn{1}{c}{Target Sources} \\
& Code & Time & VLBA & \\
& & (hours) & Antennas$^{a}$ & \\ \tableline \\[-5pt]
2013 Aug 16 & S6117D1 & 6 & FD,LA & SHBL~J001355.9$-$185406, 1ES~0033+595 \\ [3pt]
2013 Aug 23 & S6117A1 & 8 & None  & RGB~J0152+017, 1ES~0229+200, \\ [2pt]
            &         &   &       & RBS~0413, 1ES~0347$-$121 \\ [3pt]
2013 Aug 30 & S6117D2 & 6 & None  & KUV~00311$-$1938, RGB~J0136+391\\ [3pt]
2013 Sep 19 & S6117B1 & 8 & KP    & 1ES~0414+009, 1ES~0502+675, \\ [2pt]
            &         &   &       & PKS~0548$-$322, RGB~J0710+591 \\ [3pt]
2013 Oct 21 & S6117D3 & 6 & LA    & RX~J0648.7+1516, 1ES~0647+250 \\ [3pt]
2013 Oct 24 & S6117D4 & 6 & FD,LA & 1RXS~J101015.9$-$311909, MS~1221.8+2452 \\ [3pt]
2013 Dec 23 & S6117D5 & 9 & KP,NL & 1ES~1440+122, H1722+119, \\ [2pt]
            &         &   &       & 1ES~1741+196, B3~2247+381 \\ \tableline \\[-10pt]
\end{tabular}}
\end{center}
{\bf Notes.}\\
$a$: VLBA antennas that did not participate or that were excluded from the imaging 
for that session. FD=Fort Davis, Texas, KP=Kitt Peak, Arizona, LA=Los Alamos, New Mexico,
NL=North Liberty, Iowa.\\
\end{table*}

We used the AIPS software package for calibration and fringe-fitting of the correlated visibilities,
and fringes were found across the full bandwidth
at significant SNR and small delays and rates to all target sources. 
A small number of discrepant visibilities were flagged, 
and the final images were produced using CLEAN and self-calibration in
the DIFMAP software package. 
VLBA imaging of sources at these lower flux density levels can be sensitive
to the self-calibration averaging interval,
and self-calibration will generate spurious point-source structure if the averaging interval
is too short (e.g., Mart{\'{\i}}-Vidal \& Marcaide 2008).
We carefully investigated and selected self-calibration solution intervals for the fainter
sources to make sure that minimal spurious flux density (less than $\sim1$~mJy) 
should be introduced into the images through self-calibration
(see Equations~7 and 8 of Mart{\'{\i}}-Vidal \& Marcaide 2008).
In the section below, all images are displayed using natural weighting, 
in order to maximize the dynamic range.
At a typical redshift for these sources of $z\sim0.2$, 1~milliarcsecond (a typical beam size)
corresponds to a linear resolution of about 3~parsecs,
and the smallest sizes measurable in model fitting (about 10\% of the beam size)
would have a linear size of about 0.3~parsecs. 

\section{Results for the 20 New Sources}
\label{results20}

\subsection{Images}
\label{images}
\begin{figure*}[!t]
\centering
\includegraphics[scale=0.85]{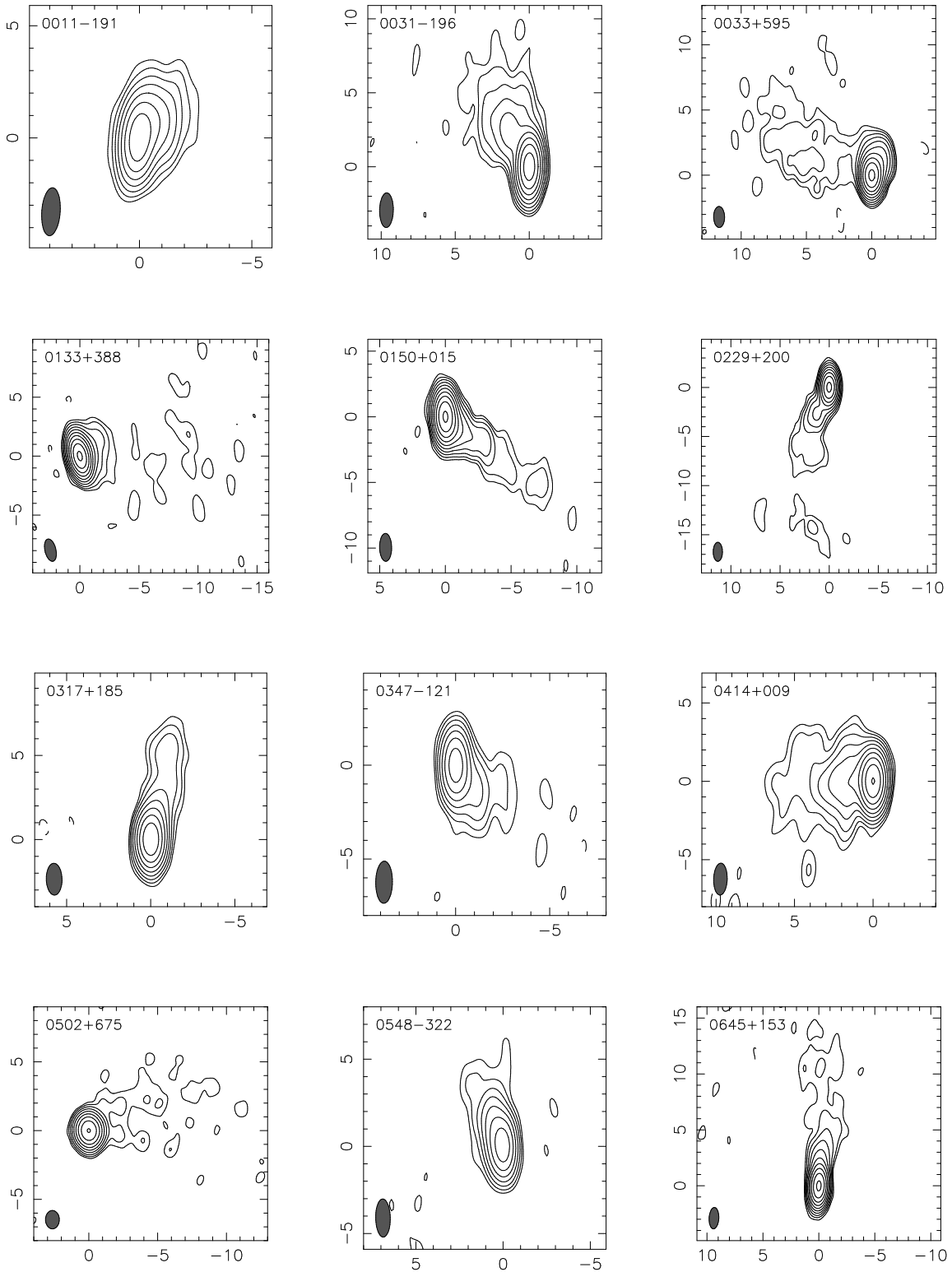}
\caption{VLBA images at 8.4~GHz of TeV blazars from Table~\ref{imtab}.
Parameters of the images are given in Table~\ref{imtab}. Axes are in milliarcseconds.
The lowest contour in each images is 3 times the rms noise level from
Table~\ref{imtab}, and each subsequent contour is a factor of two higher.}
\end{figure*}

\begin{figure*}[!t]
\centering
\includegraphics[scale=0.85]{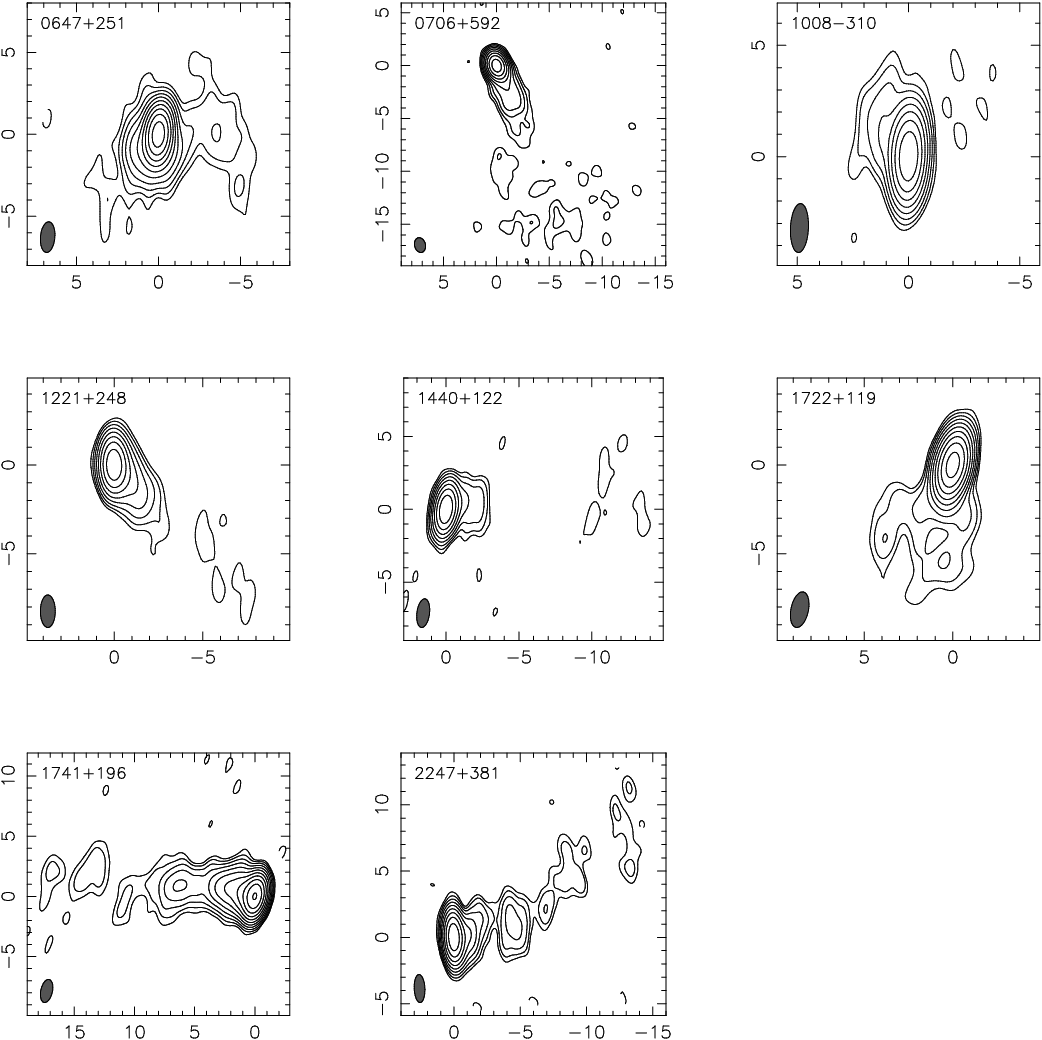}
\vspace*{0.125in}
\\Fig. 1.--—{\em Continued}
\end{figure*}

The VLBA images of the 20 TeV HBLs studied for this paper are shown in Figure~1, and the
parameters of these images are tabulated in Table~\ref{imtab}.
The B1950 name is shown in each panel in Figure~1, and may be used subsequently to refer to the source.
All sources show a bright, compact component, hereafter identified as the VLBI core,
and they all show additional extended structure that can be modeled by at least one
Gaussian feature in addition to the core (see $\S$~\ref{mfits}).
Thus, all of these sources are suitable for continued VLBI monitoring to study
the parsec-scale jet kinematics.
The images in Figure~1 do not show the entire CLEANed region for clarity, but instead are zoomed in on
the core and the inner jet region.
Larger scale images plus all associated data files are available at the project web site
\footnote{www.whittier.edu/facultypages/gpiner/research/archive/\\archive.html}.
The peak flux densities in the images in Figure~1 range from 7 to 98 mJy~bm$^{-1}$
(see Table~\ref{imtab}). However, the noise levels are quite low, typically
only about 0.02 mJy~bm$^{-1}$ (Table~\ref{imtab}), close to the expected thermal noise limit for these observations, 
so that even the images of the fainter sources have dynamic ranges of several hundred,
which is easily high enough to image the parsec-scale jet structure.

\begin{table*}[!t]
\begin{center}
{\normalsize \caption{Parameters of the Images}
\label{imtab}
\begin{tabular}{l c c c c c} \tableline \tableline \\ [-5pt]
\multicolumn{1}{c}{Source} & B1950 & Time On & Beam & Peak Flux & $I_{\rm {rms}}^{b}$ \\
& Name & Source & Parameters$^{a}$ & Density & (mJy bm$^{-1}$) \\
& & (minutes) & & (mJy bm$^{-1}$) \\ \tableline \\ [-5pt]
SHBL~J001355.9$-$185406 & 0011$-$191 & 120 & 2.15,0.82,$-$4.1  & 10 & 0.029 \\ [2pt]
KUV~00311$-$1938        & 0031$-$196 & 144 & 2.34,0.93,$-$1.1  & 26 & 0.022 \\ [2pt]
1ES~0033+595            & 0033+595   & 132 & 1.61,0.84,0.7     & 43 & 0.024 \\ [2pt]
RGB~J0136+391           & 0133+388   & 150 & 1.93,0.93,13.3    & 35 & 0.020 \\ [2pt]
RGB~J0152+017           & 0150+015   & 96  & 2.12,0.92,0.9     & 43 & 0.025 \\ [2pt]
1ES~0229+200            & 0229+200   & 96  & 1.93,0.94,$-$0.2  & 21 & 0.023 \\ [2pt]
RBS~0413                & 0317+185   & 96  & 1.89,0.94,1.3     & 18 & 0.025 \\ [2pt]
1ES~0347$-$121          & 0347$-$121 & 96  & 2.25,0.89,$-$0.9  & 7  & 0.025 \\ [2pt]
1ES~0414+009            & 0414+009   & 104 & 2.04,0.87,$-$1.7  & 35 & 0.022 \\ [2pt]
1ES~0502+675            & 0502+675   & 104 & 1.34,1.01,0.5     & 19 & 0.023 \\ [2pt]
PKS~0548$-$322          & 0548$-$322 & 104 & 2.19,0.84,1.0     & 20 & 0.062 \\ [2pt]
RX~J0648.7+1516         & 0645+153   & 160 & 1.92,0.86,$-$3.0  & 36 & 0.020 \\ [2pt]
1ES~0647+250            & 0647+251   & 160 & 1.88,0.88,$-$4.9  & 43 & 0.018 \\ [2pt]
RGB~J0710+591           & 0706+592   & 104 & 1.42,1.03,14.5    & 28 & 0.023 \\ [2pt]
1RXS~J101015.9$-$311909 & 1008$-$310 & 120 & 2.20,0.81,$-$2.7  & 29 & 0.040 \\ [2pt]
MS~1221.8+2452          & 1221+248   & 132 & 1.83,0.84,$-$0.1  & 16 & 0.023 \\ [2pt]
1ES~1440+122            & 1440+122   & 117 & 1.98,0.87,$-$7.2  & 18 & 0.025 \\ [2pt]
H~1722+119              & 1722+119   & 117 & 2.04,0.98,$-$11.8 & 66 & 0.030 \\ [2pt]
1ES~1741+196            & 1741+196   & 117 & 1.95,0.97,$-$12.6 & 98 & 0.030 \\ [2pt]
B3~2247+381             & 2247+381   & 117 & 2.11,0.81,2.5     & 42 & 0.029 \\ \tableline \\ [-10pt]
\end{tabular}}
\end{center}
{\bf Notes.}\\
$a$: Numbers are for the naturally weighted beam, and are the FWHMs of the major
and minor axes in mas, and the position angle of the major axis in degrees.
Position angle is measured from north through east.\\
$b$: Rms noise in the total intensity image.\\
\end{table*}

About half of these sources have been previously imaged with the VLBA by other investigators, although
all of those images were obtained prior to the VLBA sensitivity upgrade.
Rector et al.\ (2003) show 5~GHz VLBA images of the five sources 0033+595, 0229+200, 0414+009, 0647+251,
and 1741+196. Those images have about twice the beam size and about three times the noise level
of the images in Figure~1, but they all agree in showing the same general extended jet structure.
Giroletti et al.\ (2004b) show 5~GHz VLBA images of the five sources 0229+200, 0347$-$121, 0548$-$322, 0706+592, and
1440+122; however, those images all have about six times the noise level of the images in Figure~1, 
and they only detect parsec-scale jet structure in 0706+592.
The source 1722+119 has a single image in the MOJAVE database, but it shows only the VLBI core.
Collectively, these prior imaging results for those ten sources 
demonstrate the importance of the VLBA sensitivity upgrade in imaging the parsec-scale structure of the TeV HBLs.
For the remaining ten sources in Figure~1, these are the first published VLBI images known to the authors.

The general parsec-scale morphology of the sources in Figure~1 is familiar from VLBI studies
of brighter TeV blazars; for example, Mrk~421 and Mrk~501 
(Piner et al.\ 1999; Edwards \& Piner 2002; Giroletti et al.\ 2006, 2008).
Most of the sources show a collimated jet a few milliarcseconds long that transitions
to a lower surface brightness, more diffuse jet with a broader opening angle at a few mas from the core.
The structure at tens of milliarcseconds from the core at 8~GHz then appears patchy and filamentary.
As an example, the source 0706+592 nicely displays this morphology in Figure~1.
Despite this general pattern, there are a couple of sources with unusual morphologies.
The sources 0033+595 and 0647+251 both show structure on opposite sides of the 
presumed core. Either the brightest most compact component is not the core,
the jet crosses back over the line of sight (as seen in the TeV blazar 1ES~1959+650 by Piner et al.\ 2008),
or the emission is truly two-sided. Forthcoming imaging at multiple frequencies should identify the core for these unusual cases.
At least two sources (0502+675 and 1722+119) also display limb-brightened jets in 
their inner jet region, which is discussed further in $\S$~\ref{transverse}.

\subsection{Model Fits}
\label{mfits}
After imaging and final calibration of the visibilities, 
we fit circular Gaussian models to the calibrated visibilities for each source using the
{\em modelfit} task in DIFMAP. Circular Gaussians are more stable than elliptical Gaussians during fitting,
and they provided adequate fits to the visibilities for all sources,
as noted by the reduced chi-squared of the fit and visual inspection of the residual map and visibilities.
Model fitting directly to the visibilities allows sub-beam
resolution to be obtained, and components can be clearly identified in the model fitting even when they
appear blended with the core component or with each other in the CLEAN images.
In a number of cases patchy low surface brightness emission beyond the collimated jet region
could not be well-fit by a circular Gaussian,
so the model fits do not necessarily represent the most distant emission seen on the CLEAN images.
Note also that, because of incomplete sampling in the $(u,v)$-plane,
VLBI model fits are not unique, and represent
only one mathematically possible deconvolution of the source structure.

The circular Gaussian models fit to all 20 sources are given in Table~\ref{mfittab}.
Note that flux values for closely spaced components may be inaccurate, since it is difficult for
the fitting algorithm to uniquely distribute the flux during model fitting.
The model component naming follows the scheme used in our previous papers (e.g., Piner et al.\ 2010);
jet components are numbered C1, C2, etc., from the outermost component inward.
Observer-frame brightness temperatures are also given in Table~\ref{mfittab} for all 
partially-resolved core components (those whose best-fit size is not zero).
These VLBI core brightness temperatures and associated errors are discussed in detail in the following subsection.

\begin{table*} [!h]
\begin{center}
{\normalsize \caption{Circular Gaussian Models}
\label{mfittab}
\begin{tabular}{l c c r r r c c c} \tableline \tableline \\ [-5pt]
\multicolumn{1}{c}{Source} & B1950 & Component & \multicolumn{1}{c}{$S$} & \multicolumn{1}{c}{$r$} &
\multicolumn{1}{c}{PA} & $a$ & $\chi_{R}^{2}$ & $T_{B}$ \\
\multicolumn{1}{c}{(1)} & Name & (3) & \multicolumn{1}{c}{(mJy)} & \multicolumn{1}{c}{(mas)} &
\multicolumn{1}{c}{(deg)} & (mas) & (8) & ($10^{10}$ K) \\
& (2) & & \multicolumn{1}{c}{(4)} & \multicolumn{1}{c}{(5)} & \multicolumn{1}{c}{(6)} & (7) & & (9) \\ \tableline \\ [-5pt]
SHBL~J001355.9$-$185406 & 0011$-$191 & Core &  9.3 & ...   & ...      & 0.23  & 0.66 & 0.3 \\ [2pt]
                        &            & C1   &  5.0 & 0.99  & $-$42.4  & 0.98  &      &     \\ [2pt]
KUV~00311$-$1938        & 0031$-$196 & Core & 25.8 & ...   & ...      & 0.10  & 0.74 & 4.3 \\ [2pt]
                        &            & C2   &  2.7 & 2.55  & 29.9     & 1.39  &      &     \\ [2pt]
                        &            & C1   &  2.6 & 4.56  & 22.7     & 3.54  &      &     \\ [2pt]
1ES~0033+595            & 0033+595   & Core & 43.7 & ...   & ...      & 0.23  & 0.72 & 1.5 \\ [2pt]
                        &            & C2   &  9.7 & 1.18  & $-$24.1  & 1.16  &      &     \\ [2pt]
                        &            & C1   &  5.1 & 5.52  & 69.1     & 4.30  &      &     \\ [2pt]
RGB~J0136+391           & 0133+388   & Core & 28.7 & ...   & ...      & 0.00  & 0.75 & ... \\ [2pt]
                        &            & C3   &  6.8 & 0.36  & $-$7.9   & 0.52  &      &     \\ [2pt]
                        &            & C2   &  2.3 & 0.97  & $-$59.6  & 1.92  &      &     \\ [2pt]
                        &            & C1   &  2.5 & 7.99  & $-$92.4  & 9.45  &      &     \\ [2pt]
RGB~J0152+017           & 0150+015   & Core & 42.6 & ...   & ...      & 0.16  & 0.80 & 2.9 \\ [2pt]
                        &            & C2   &  4.8 & 1.02  & $-$135.6 & 0.68  &      &     \\ [2pt]
                        &            & C1   &  2.9 & 3.27  & $-$125.4 & 1.61  &      &     \\ [2pt]
1ES~0229+200            & 0229+200   & Core & 19.9 & ...   & ...      & 0.10  & 0.74 & 3.6 \\ [2pt]
                        &            & C4   &  2.2 & 0.94  & 163.6    & 0.50  &      &     \\ [2pt]
                        &            & C3   &  2.3 & 3.06  & 154.6    & 1.04  &      &     \\ [2pt]
                        &            & C2   &  1.4 & 6.71  & 160.8    & 2.80  &      &     \\ [2pt]
                        &            & C1   &  2.4 & 15.76 & 171.6    & 8.79  &      &     \\ [2pt]
RBS~0413                & 0317+185   & Core & 16.7 & ...   & ...      & 0.08  & 0.74 & 4.5 \\ [2pt]
                        &            & C3   &  2.6 & 0.85  & $-$13.9  & 0.38  &      &     \\ [2pt]
                        &            & C2   &  1.1 & 2.12  & $-$14.8  & 0.70  &      &     \\ [2pt]
                        &            & C1   &  1.2 & 5.03  & $-$11.2  & 1.53  &      &     \\ [2pt]
1ES~0347$-$121          & 0347$-$121 & Core &  7.4 & ...   & ...      & 0.14  & 0.89 & 0.6 \\ [2pt]
                        &            & C1   &  1.6 & 1.74  & $-$144.2 & 1.38  &      &     \\ [2pt]
1ES~0414+009            & 0414+009   & Core & 35.7 & ...   & ...      & 0.28  & 0.78 & 0.8 \\ [2pt]
                        &            & C1   & 11.1 & 1.41  & 85.7     & 2.59  &      &     \\ [2pt]
1ES~0502+675            & 0502+675   & Core & 17.2 & ...   & ...      & 0.26  & 0.75 & 0.4 \\ [2pt]
                        &            & C2   &  2.9 & 0.37  & $-$139.0 & 0.41  &      &     \\ [2pt]
                        &            & C1   &  3.2 & 4.58  & $-$74.5  & 6.90  &      &     \\ [2pt]
PKS~0548$-$322          & 0548$-$322 & Core & 20.4 & ...   & ...      & 0.32  & 0.69 & 0.3 \\ [2pt]
                        &            & C1   &  6.2 & 1.27  & 30.9     & 0.61  &      &     \\ [2pt]
RX~J0648.7+1516         & 0645+153   & Core & 33.7 & ...   & ...      & 0.00  & 0.76 & ... \\ [2pt]
                        &            & C5   &  3.9 & 0.81  & $-$1.9   & 0.32  &      &     \\ [2pt]
                        &            & C4   &  2.1 & 2.43  & $-$1.6   & 0.87  &      &     \\ [2pt]
                        &            & C3   &  1.1 & 4.67  & $-$10.9  & 2.08  &      &     \\ [2pt]
                        &            & C2   &  2.6 & 10.77 & $-$1.3   & 5.09  &      &     \\ [2pt]
                        &            & C1   &  3.3 & 21.51 & 5.6      & 7.18  &      &     \\ [2pt]
1ES~0647+250            & 0647+251   & Core & 41.6 & ...   & ...      & 0.15  & 0.76 & 3.2 \\ [2pt]
                        &            & C2   &  9.1 & 0.91  & 157.9    & 1.78  &      &     \\ [2pt]
                        &            & C1   &  1.8 & 3.28  & $-$82.6  & 3.29  &      &     \\ [2pt]
RGB~J0710+591           & 0706+592   & Core & 27.0 & ...   & ...      & 0.17  & 0.75 & 1.5 \\ [2pt]
                        &            & C3   &  4.8 & 0.80  & $-$145.3 & 0.63  &      &     \\ [2pt]
                        &            & C2   &  3.8 & 3.10  & $-$156.0 & 1.83  &      &     \\ [2pt]
                        &            & C1   &  4.9 & 14.50 & $-$156.7 & 10.57 &      &     \\ [2pt]
\end{tabular}}
\end{center}
\end{table*}

\begin{table*}[!t]
\begin{center}
{\normalsize Table 4 (cont.): Circular Gaussian Models \\
\begin{tabular}{l c c r r r c c c} \tableline \tableline \\ [-5pt]
\multicolumn{1}{c}{Source} & B1950 & Component & \multicolumn{1}{c}{$S$} & \multicolumn{1}{c}{$r$} &
\multicolumn{1}{c}{PA} & $a$ & $\chi_{R}^{2}$ & $T_{B}$ \\
\multicolumn{1}{c}{(1)} & Name & (3) & \multicolumn{1}{c}{(mJy)} & \multicolumn{1}{c}{(mas)} &
\multicolumn{1}{c}{(deg)} & (mas) & (8) & ($10^{10}$ K) \\
& (2) & & \multicolumn{1}{c}{(4)} & \multicolumn{1}{c}{(5)} & \multicolumn{1}{c}{(6)} & (7) & & (9) \\ \tableline \\ [-5pt]
1RXS~J101015.9$-$311909 & 1008$-$310 & Core & 30.1 & ...   & ...      & 0.29  & 0.65 & 0.6 \\ [2pt]
                        &            & C1   &  5.1 & 0.91  & 32.7     & 2.01  &      &     \\ [2pt]
MS~1221.8+2452          & 1221+248   & Core & 16.0 & ...   & ...      & 0.22  & 0.68 & 0.6 \\ [2pt]
                        &            & C2   &  2.6 & 0.77  & $-$129.5 & 0.65  &      &     \\ [2pt]
                        &            & C1   &  1.6 & 2.15  & $-$139.0 & 1.12  &      &     \\ [2pt]
1ES~1440+122            & 1440+122   & Core & 16.7 & ...   & ...      & 0.00  & 0.67 & ... \\ [2pt]
                        &            & C3   &  2.5 & 0.48  & $-$49.6  & 0.42  &      &     \\ [2pt]
                        &            & C2   &  1.5 & 1.68  & $-$71.8  & 1.47  &      &     \\ [2pt]
                        &            & C1   &  1.3 & 12.41 & $-$81.6  & 6.59  &      &     \\ [2pt]
H~1722+119              & 1722+119   & Core & 62.5 & ...   & ...      & 0.00  & 0.68 & ... \\ [2pt]
                        &            & C2   &  3.7 & 0.58  & 152.4    & 0.00  &      &     \\ [2pt]
                        &            & C1   &  4.0 & 3.50  & 163.3    & 4.29  &      &     \\ [2pt]
1ES~1741+196            & 1741+196   & Core & 94.8 & ...   & ...      & 0.24  & 0.78 & 2.9 \\ [2pt]
                        &            & C3   & 22.0 & 0.75  & 70.0     & 0.67  &      &     \\ [2pt]
                        &            & C2   & 17.5 & 2.08  & 71.7     & 1.50  &      &     \\ [2pt]
                        &            & C1   &  7.0 & 6.00  & 81.1     & 2.43  &      &     \\ [2pt]
B3~2247+381             & 2247+381   & Core & 39.8 & ...   & ...      & 0.16  & 0.75 & 2.7 \\ [2pt]
                        &            & C3   & 10.3 & 0.69  & $-$91.5  & 1.53  &      &     \\ [2pt]
                        &            & C2   &  3.4 & 4.63  & $-$74.6  & 1.70  &      &     \\ [2pt]
                        &            & C1   &  1.7 & 9.61  & $-$59.7  & 3.65  &      &     \\ \tableline \\ [-10pt]
\end{tabular}}
\end{center}
{\bf Notes.}
Column~4: flux density in millijanskys.
Columns~5 and 6: $r$ and PA are the polar coordinates of the
center of the component relative to the presumed core.
Position angle is measured from north through east.
Column~7: FWHM of the Gaussian component.
Column~8: the reduced chi-squared of the model fit.
Column~9: the maximum observer-frame brightness temperature of the Gaussian core component is given by
$T_{B}=1.22\times10^{12}S/(a^{2}\nu^{2})$~K,
where $S$ is the flux density in janskys,
$a$ is the FWHM in mas, and $\nu$ is the observation frequency in GHz.
Brightness temperature is given for core components whose best-fit size is not zero.
\end{table*}

\subsection{Core Brightness Temperatures}
\label{tb}
The observed brightness temperatures of VLBI cores can be used to constrain
both Doppler beaming factors and the physical processes occurring in a source.
The maximum observer-frame brightness temperature of a circular Gaussian is 
\begin{equation}
\label{tbeq}
T_{B}=1.22\times10^{12}\;\frac{S}{a^{2}\nu^{2}}\;\rm{K},
\end{equation}
where $S$ is the flux density of the Gaussian in janskys, $a$ is the FWHM of the Gaussian in mas, and
$\nu$ is the observing frequency in GHz.
Note that we use observer-frame brightness temperatures, i.e., without applying the $(1+z)$ factor
to convert to source-frame brightness temperatures, because the redshift of a number of
these sources is uncertain (see Table~\ref{alltab}). The median redshift of the sources with known redshift is
about 0.2, so source-frame brightness temperatures are only about 20\% higher.

Several mechanisms can act to limit the intrinsic rest-frame brightness temperature of a synchrotron source; e.g.,
rapid energy loss by inverse Compton emission that limits the
brightness temperature to $\sim5\times10^{11} - 1\times10^{12}$~K (Kellermann \& Pauliny-Toth 1969), or
equipartition of energy between particles and magnetic fields that limits the brightness temperature to
$\sim5\times10^{10} - 1\times10^{11}$~K (Readhead 1994).
The observed brightness temperature of a source is increased relative to its intrinsic
brightness temperature by a factor of the Doppler factor $\delta$
\footnote{The Doppler factor $\delta=1/(\gamma(1-\beta\cos\theta))$,
where $\theta$ is the viewing angle, $\beta =v/c$,
and $\gamma=(1-\beta^2)^{-1/2}$ is the bulk Lorentz factor.}.
Thus, if upper limits to the intrinsic brightness temperature are known, then the
observed brightness temperature can be used to compute a lower limit to the Doppler factor.
Similarly, if the Doppler factor can be estimated by independent means, then intrinsic brightness
temperatures can be computed from observed values. This has been done by, e.g., 
L\"{a}hteenm\"{a}ki et al.\ (1999), Homan et al.\ (2006), and Hovatta et al.\ (2013), who all used
either apparent superluminal speeds or total flux density variability
to compute intrinsic brightness temperatures of AGN samples. All of these studies concluded that
typical intrinsic brightness temperatures were in the range of few times $10^{10}$ to $10^{11}$~K,
and therefore likely to be limited by equipartition of energy. 

Care must be taken in the analysis of these VLBI core brightness temperatures, because many of the cores
are only partially resolved, and even though a `best-fit' size may be returned by the model fitting routine,
the fit is often nearly as good if the Gaussian component is simply replaced by a delta function.
In such cases, only an upper limit to the size, or a lower limit to the brightness temperature,
can actually be measured. The flux density of the core, and the baseline lengths and sensitivity of the VLBI array,
determine the maximum measurable brightness temperature (e.g., Lovell et al.\ 2000; Wehrle et al.\ 2001;
Kovalev et al.\ 2005; Lobanov 2005), which is of order $10^{11}$~K for these observations.
Because some core brightness temperatures in Table~\ref{mfittab} are within factors of a few
of this value, we conducted a full error analysis of the core brightness temperatures; this 
error analysis is described below and tabulated in Table~\ref{tbtab}.

\begin{table*}[!t]
\begin{center}
{\normalsize \caption{Brightness Temperature Error Analysis}
\label{tbtab}
\begin{tabular}{c r c c r c c r c c} \tableline \tableline  \\ [-5pt]
B1950 & \multicolumn{1}{c}{$S$} & $a$ & $T_{B}$ 
& \multicolumn{1}{c}{$S_{\rm{max}}$} & $a_{\rm{min}}$ & $T_{B,\rm{max}}$ 
& \multicolumn{1}{c}{$S_{\rm{min}}$} & $a_{\rm{max}}$ & $T_{B,\rm{min}}$ \\ 
Name & \multicolumn{1}{c}{(mJy)} & (mas) & ($10^{10}$ K) & \multicolumn{1}{c}{(mJy)} & (mas) & ($10^{10}$ K)
& \multicolumn{1}{c}{(mJy)} & (mas) & ($10^{10}$ K) \\
(1) & \multicolumn{1}{c}{(2)} & (3) & (4) & \multicolumn{1}{c}{(5)} & (6) & (7) & \multicolumn{1}{c}{(8)} & (9) & (10) \\ \tableline \\ [-5pt]
0011$-$191 & 9.3  & 0.23 & 0.3      & 13.4 & 0.00	& $\infty$ & 7.1  & 0.36 & 0.1 \\ [2pt]
0031$-$196 & 25.8 & 0.10 & 4.3      & 29.3 & 0.00	& $\infty$ & 22.3 & 0.19 & 1.1 \\ [2pt]
0033+595   & 43.7 & 0.23 & 1.5      & 48.2 & 0.14	& 4.4      & 39.2 & 0.30 & 0.8 \\ [2pt]
0133+388   & 28.7 & 0.00 & $\infty$ & 33.3 & 0.00	& $\infty$ & 22.3 & 0.27 & 0.5 \\ [2pt]
0150+015   & 42.6 & 0.16 & 2.9      & 45.1 & 0.09	& 10.3     & 38.1 & 0.23 & 1.3 \\ [2pt]
0229+200   & 19.9 & 0.10 & 3.6      & 23.4 & 0.00	& $\infty$ & 16.4 & 0.23 & 0.5 \\ [2pt]
0317+185   & 16.7 & 0.08 & 4.5      & 19.2 & 0.00	& $\infty$ & 14.2 & 0.22 & 0.5 \\ [2pt]
0347$-$121 & 7.4  & 0.14 & 0.6      & 10.9 & 0.00	& $\infty$ & 4.9  & 0.28 & 0.1 \\ [2pt]
0414+009   & 35.7 & 0.28 & 0.8      & 40.2 & 0.17	& 2.4      & 32.2 & 0.33 & 0.5 \\ [2pt]
0502+675   & 17.2 & 0.26 & 0.4      & 22.7 & 0.11	& 3.1      & 13.7 & 0.37 & 0.2 \\ [2pt]
0548$-$322 & 20.4 & 0.32 & 0.3      & 26.9 & 0.12	& 3.0      & 15.9 & 0.36 & 0.2 \\ [2pt]
0645+153   & 33.7 & 0.00 & $\infty$ & 39.0 & 0.00	& $\infty$ & 30.0 & 0.21 & 1.2 \\ [2pt]
0647+251   & 41.6 & 0.15 & 3.2      & 45.1 & 0.10	& 8.2      & 38.1 & 0.23 & 1.2 \\ [2pt]
0706+592   & 27.0 & 0.17 & 1.5      & 31.5 & 0.00	& $\infty$ & 22.5 & 0.29 & 0.5 \\ [2pt]
1008$-$310 & 30.1 & 0.29 & 0.6      & 37.6 & 0.13	& 3.9      & 24.6 & 0.39 & 0.3 \\ [2pt]
1221+248   & 16.0 & 0.22 & 0.6      & 20.5 & 0.00	& $\infty$ & 12.5 & 0.34 & 0.2 \\ [2pt]
1440+122   & 16.7 & 0.00 & $\infty$ & 20.2 & 0.00	& $\infty$ & 13.2 & 0.23 & 0.4 \\ [2pt]
1722+119   & 62.5 & 0.00 & $\infty$ & 65.9 & 0.00	& $\infty$ & 58.9 & 0.13 & 6.0 \\ [2pt]
1741+196   & 94.8 & 0.24 & 2.9      & 98.0 & 0.21	& 3.8      & 89.2 & 0.28 & 1.9 \\ [2pt]
2247+381   & 39.8 & 0.16 & 2.7      & 44.3 & 0.00	& $\infty$ & 35.3 & 0.23 & 1.1 \\ \tableline \\ [-10pt]
\end{tabular}}
\end{center}
{\bf Notes.}
Columns 2, 3, and 4 are the best-fit core flux density and size, and the associated 
observer's frame brightness temperature. These values are also given in Table~\ref{mfittab}.
Columns 5, 6, and 7 are the maximum allowed core flux density and the minimum allowed size, and the associated maximum brightness
temperature computed from those quantities.
Columns 8, 9, and 10 are the minimum allowed core flux density and the maximum allowed size, and the associated minimum brightness
temperature computed from those quantities.
\end{table*}

We used the Difwrap program (Lovell 2000), as described by e.g., Piner et al.\ (2000) and Tingay et al.\ (2001),
to determine upper and lower bounds to the measured brightness temperatures.
We established minimum and maximum values for the flux density and size of a component by 
systematically varying that property, while allowing other parameters to re-converge, and then visually
comparing the new fit to the measured visibilities.
The upper bound to the brightness temperature was then computed using the maximum flux and
the minimum size, while the lower bound to the brightness temperature was computed using the minimum
flux and the maximum size. All of these values are tabulated in Table~\ref{tbtab}. 

Roughly half of the core components are consistent with a size of zero, meaning
that the associated brightness temperature measurements
have no upper bound and are only lower limits, but the half that do have upper bounds provide valuable constraints.
If we exclude the single high-brightness temperature source 1722+119 ($T_{B}>6.0\times10^{10}$~K),
then the largest lower-limit is $T_{B}>1.9\times10^{10}$~K for 1741+196. 
Similarly, the smallest upper-limit is $T_{B}<2.4\times10^{10}$~K for 0414+009.
Thus, all except one of these twenty sources are consistent with the brightness temperature range
$1.9\times10^{10}<T_{B}<2.4\times10^{10}$~K, and we therefore take a brightness temperature of
$\sim2\times10^{10}$~K to be a typical observed brightness temperature of a TeV HBL.
This is consistent with typical observed brightness temperatures of TeV HBLs measured in earlier works
(e.g., Piner et al.\ 2010), but it is now established for a much larger number of sources.
To compare with intrinsic brightness temperature limits that have been derived for homogeneous optically thick spheres, we
can convert our Gaussian brightness temperatures to homogeneous sphere brightness temperatures
by multiplying by the appropriate correction factor of 0.56 (e.g., Pearson 1995; Tingay et al.\ 2001).
This yields a value of about $1\times10^{10}$~K as a typical brightness temperature of a TeV HBL.

Comparing the typical observed brightness temperatures to the equipartition brightness temperatures
calculated for these sources of about $6\times10^{10}$~K (Equation~4$a$ of Readhead 1994),
we see that the observed brightness temperatures of these TeV HBLs are already at or below the equipartition
limit with no need to invoke high Doppler factors to reduce the observed brightness temperatures.
There is thus no evidence of relativistic beaming of the core emission based on the VLBI brightness temperatures.
Even with no Doppler boosting, the observed brightness temperatures are already somewhat
below the equipartition value, placing these sources in the magnetically-dominated regime
(Readhead 1994; Homan et al.\ 2006). High values of the Doppler factor would reduce the 
intrinsic brightness temperature even more, placing the sources even farther from equipartition.
We note one important caveat: that both observed brightness temperatures and intrinsic physical limits
are traditionally calculated for a homogeneous sphere geometry, and that the actual geometry for the VLBI
core region may be more complex, such as a partially-resolved limb-brightened structure (see $\S$~\ref{transverse}).

Despite the lack of evidence for beaming from the VLBI brightness temperatures, 
the one-sided core-jet morphology displayed by the majority of these sources does imply at least mild Doppler boosting.
However, because the sources studied in this paper are relatively faint, this constraint is modest. 
We have computed lower limits to the jet-to-counterjet brightness ratio for each source,
based on the peak jet brightness from the model fits in Table~\ref{mfittab} restored with 
the associated beam from Table~\ref{imtab}, and using three times the rms noise from Table~\ref{imtab}
as the minimum detectable counterjet brightness.  
The median lower limit to the jet-to-counterjet brightness ratio is 37:1, 
which implies $\delta>2$ for viewing angles of a few degrees.
The highest lower limit is 210:1 for 1741+196, which implies $\gamma>2$ and $\delta>4$
for viewing angles of a few degrees.

\subsection{Opening Angles}
\label{opening}
We have calculated the apparent opening angle $\phi_{app}$ of each of these twenty jets, using the 
model fits from Table~\ref{mfittab}, and the model-fitting approach to measuring apparent opening angles
described by Pushkarev et al.\ (2009). These apparent opening angles are tabulated in Table~\ref{optab}.
Note that the apparent opening angle is a function of both the intrinsic opening angle and
the viewing angle through the relation $\phi_{app}\approx\phi_{int}/\sin\theta$, where $\theta$ is the viewing angle.
Pushkarev et al.\ (2009) compared apparent opening angles of {\em Fermi}-detected and non-detected
blazars, and found a tendency for the {\em Fermi}-detected blazars to have wider apparent
opening angles than the non-detected ones. Because the calculated intrinsic opening angles of the
two groups were similar, they suggested that the {\em Fermi}-detected jets were viewed more closely
to the line of sight. Lister et al.\ (2011) used a larger sample of {\em Fermi}-detected blazars
to compute a mean apparent opening angle of 24$\arcdeg$ for this sample, and they also found a positive correlation
between apparent opening angle and the gamma-ray loudness of the source.

\begin{table*}
\begin{center}
{\normalsize \caption{Apparent Opening Angles}
\label{optab}
\begin{tabular}{c c c c c c c c} \tableline \tableline \\ [-5pt] 
B1950 & $\phi_{app}$ & B1950 & $\phi_{app}$ & B1950 & $\phi_{app}$ & B1950 & $\phi_{app}$ \\
Name & (deg) & Name & (deg) & Name & (deg) & Name & (deg)\\ \tableline \\ [-5pt]
0011$-$191 & 26.2 & 0229+200   & 13.0 & 0548$-$322 & 13.4 & 1221+248   & 18.9 \\ [2pt]
0031$-$196 & 18.2 & 0317+185   & 10.2 & 0645+153   & 11.3 & 1440+122   & 20.7 \\ [2pt]
0033+595   & 23.8 & 0347$-$121 & 21.6 & 0647+251   & 35.5 & 1722+119   & 31.5 \\ [2pt]
0133+388   & 37.2 & 0414+009   & 42.6 & 0706+592   & 19.3 & 1741+196   & 18.4 \\ [2pt]
0150+015   & 16.2 & 0502+675   & 33.1 & 1008$-$310 & 47.8 & 2247+381   & 23.0 \\ \tableline \\ [-10pt]
\end{tabular}}
\end{center}
\end{table*}

The apparent opening angles in Table~\ref{optab} range from 10$\arcdeg$ to 48$\arcdeg$, with a 
mean of $24\pm2\arcdeg$, identical to the mean found by Lister et al.\ (2011) for a
larger sample of {\em Fermi}-detected blazars. Because  16 out of 20 of the TeV sources imaged
for this paper are also {\em Fermi} sources (see Table~\ref{alltab}), this identical mean is not surprising,
but it does show that the apparent opening angle distribution for a TeV HBL-selected subset of {\em Fermi} sources is
similar to the overall {\em Fermi}-detected distribution.
There is thus no evidence from apparent opening angles that TeV HBLs have
different distributions of either viewing angle or intrinsic opening angle compared to the larger sample
of Fermi sources studied by Lister et al.\ (2011).
We find no significant correlation between the apparent opening angles in Table~\ref{optab} and TeV gamma-ray loudness
(see $\S$~\ref{loudness}) as was found by Lister et al.\ (2011); however, this is not conclusive
considering our small sample size of 20 sources. 

\subsection{Morphology and Transverse Jet Structure}
\label{transverse}
\begin{figure*}[!t]
\centering
\includegraphics[scale=0.65]{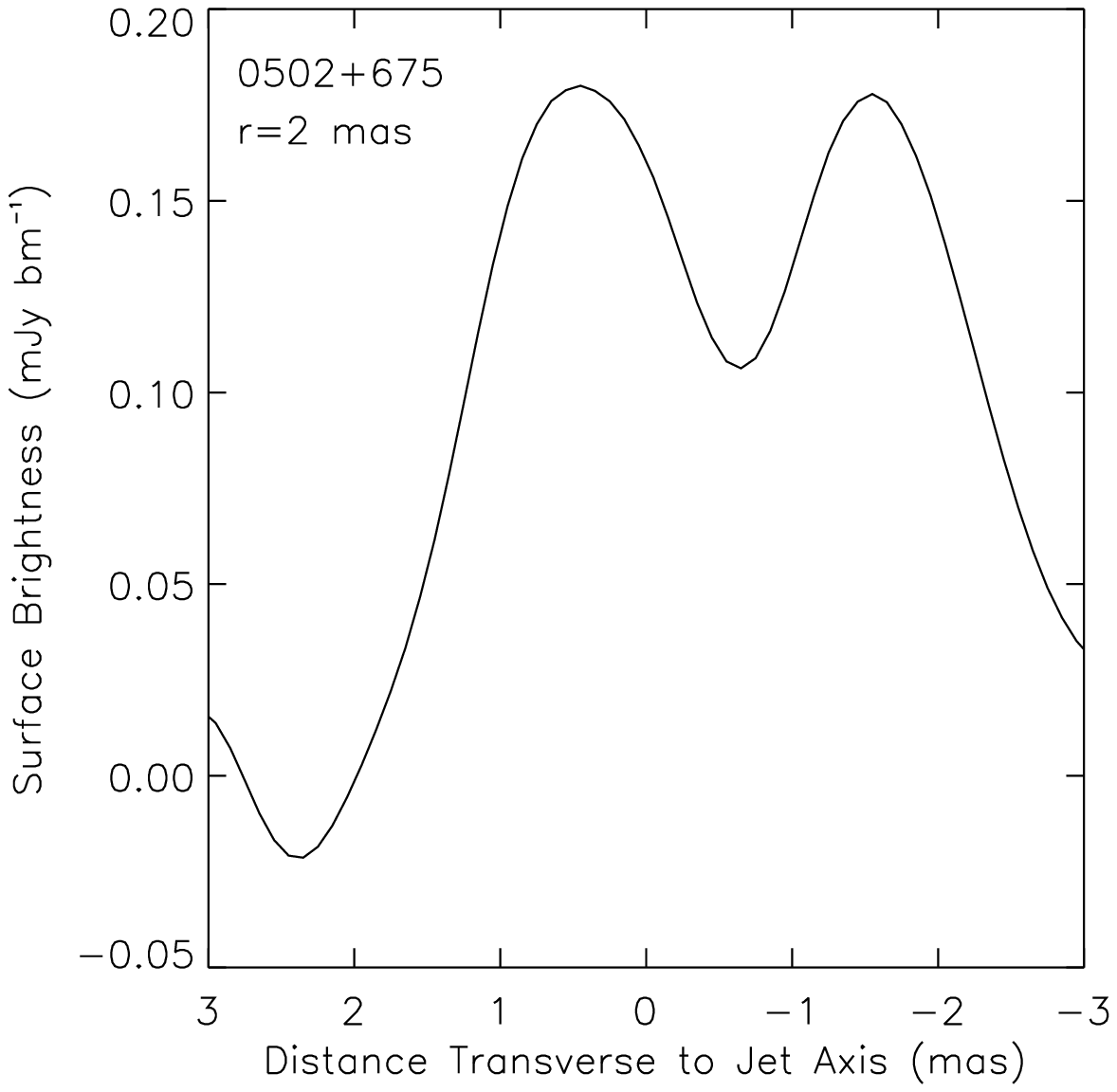}
\includegraphics[scale=0.65]{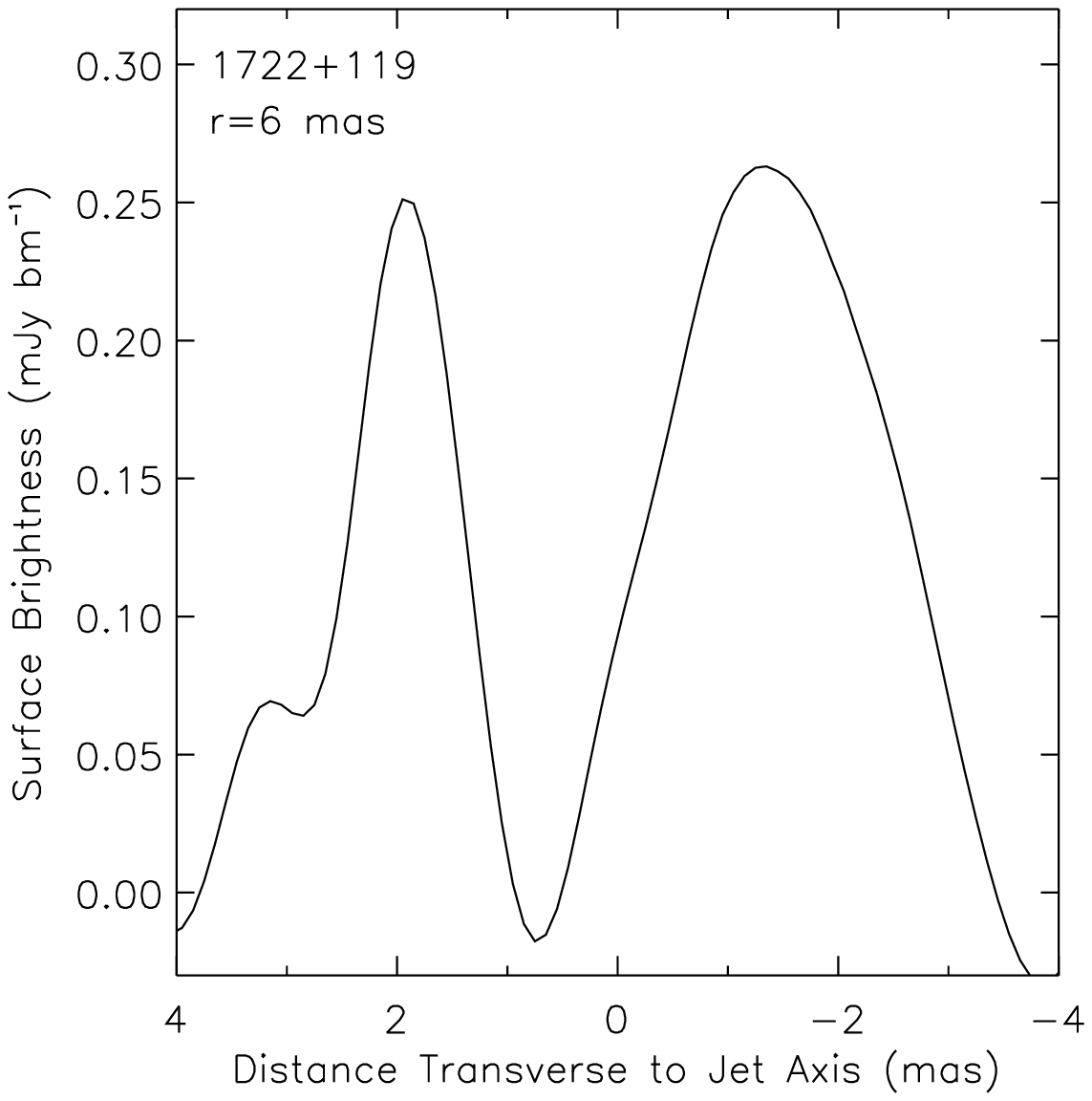}
\caption{Transverse brightness profiles showing limb-brightening for 0502+675 at 2~mas from the core (left panel), and 
for 1722+119 at 6~mas from the core (right panel).
The rms noise levels (or uncertainty on the curves) are 0.023~mJy~bm$^{-1}$ for 0502+675, 
and 0.030~mJy~bm$^{-1}$ for 1722+119 (see Table~\ref{imtab}).} 
\end{figure*}

AGN jets may be expected to develop transverse (so-called `spine-sheath' or `spine-layer') velocity structures
on theoretical grounds (e.g., Henri \& Pelletier 1991),
and the existence of these structures could explain some important
observed properties of the TeV HBLs. For example, Ghisellini et al.\ (2005) consider how a jet with a low Lorentz factor
layer and a high Lorentz factor spine could produce the discrepant Lorentz factors that are observed
for TeV HBLs in the radio and gamma-ray, while at the same time the interaction between these two
regions could serve to decelerate the spine. Recently, Tavecchio et al.\ (2014)
have calculated whether spine-layer structures in the TeV HBLs could also produce high-energy PeV neutrinos,
such as those detected by IceCube (Aartsen et al.\ 2014).

Such spine-sheath structures might produce limb-brightening in the VLBI images of these jets if,
for example, the jet bends away from the line-of-sight such that the low Lorentz factor
layer acquires a higher Doppler factor than the high Lorentz factor spine
(e.g., Giroletti et al.\ 2004a), or if the layer simply has a higher
synchrotron emissivity in the radio than the spine (e.g., Sahayanathan 2009; Ghisellini et al.\ 2005).
Note though that the presence of transverse intensity structures in VLBI images 
can have causes other than two-component outflows; for example,
Clausen-Brown et al.\ (2011) show that limb-brightening can be observed for a uniform cylindrical jet 
with a helical magnetic field and no transverse structure under certain viewing geometries.

Limb-brightening has been observed a number of times in VLBI images of two of the brightest and closest TeV HBLs:
Mrk~421 and Mrk~501. Giroletti et al.\ (2004a, 2008), Piner et al.\ (2009), and Croke et al.\ (2010) have all reported limb-brightening
in VLBI images of Mrk~501, at a wide variety of distances from the core.
Similar results have been obtained for Mrk~421, both at 
lower frequencies (Giroletti et al.\ 2006) and at 43~GHz (Piner et al.\ 2010; Blasi et al.\ 2013).
Transverse polarization structures, both in EVPA and fractional polarization, have also been observed in both
of these sources, as well as in the TeV HBL 1ES~1959+650 (Piner et al.\ 2010).

We have produced transverse brightness profiles for all 20 of 
the sources imaged for this paper, at numerous points along their jets.
Many of these sources display the following general pattern in their transverse structure: the jets are well collimated and unresolved in
the transverse direction for the first few milliarcseconds, after which they transition to patchy low surface
brightness emission that is resolved but has numerous intensity peaks in a transverse brightness profile.
Of the twenty sources, we see only two examples of a classic limb-brightening profile, in the sources 0502+675 and 1722+119.
Transverse brightness profiles showing the limb-brightening for these two sources are shown in Figure~2;
the limb-brightened structure of these two sources can also be seen directly on the images shown in Figure~1.
For both of these sources, the limb brightening remains visible over a radial range of roughly 3~mas.

We note that the absence of such a clear signature of limb-brightening in
the other sources does not mean that such transverse intensity structure is non-existent in these jets.
For example, the observations of limb-brightening by Piner et al.\ (2009) and Piner et al.\ (2010) in 
the relatively nearby TeV HBLs Mrk~501 and Mrk~421 were obtained from high-resolution 43~GHz observations,
and after subtraction of the core and super-resolution of the jet in the transverse direction.
That same linear scale would correspond to an angular separation that is
well within the jet region that is transversely unresolved in the lower-resolution 8~GHz images 
of the more distant TeV HBLs presented in this paper.
It is possible that the inner jets of these sources would display such structures if they could be transversely
resolved with high-frequency VLBI; unfortunately, observations at high-frequency are much less sensitive, and
all but the brightest few TeV HBLs are too faint for this. 
Whether `spine-sheath' structures are a nearly universal structure
for TeV HBL jets is therefore ambiguous from these observations.

\section{Results for the Entire TeV HBL Sample}
\label{resultsall}
\subsection{VLBI and Gamma-Ray Data for the Sample}
\label{dataall}
In this section, we combine the new VLBI data obtained in this paper with gamma-ray
and VLBI data on the other TeV HBLs. The VLBI and gamma-ray properties of the
44 TeV HBLs currently listed in the TeVCat catalog are tabulated in Table~\ref{alltab}.
We have attempted to quote a redshift value for every source in Table~\ref{alltab}, but in a number of cases
(9 out of 44) these values are either uncertain or they are lower limits. These cases are clearly
indicated in the notes to Table~\ref{alltab}, and those values should be used with caution. 

\begin{sidewaystable*}
\begin{center}
\vspace*{3.5in}
{\footnotesize \caption{VLBI and Gamma-Ray Properties of the TeV HBLs}
\label{alltab}
\begin{tabular}{l c c c c c c c c c c c c c c c} \tableline \tableline \\ [-4pt]
\multicolumn{1}{c}{Source} & $z$ & $\nu$ & VLBI & VLBI & $T_{B}$ & Ref & TeV & TeV & Cutoff & Ind & Ref & {\em Fermi} & Ind & Log & $\overline{m}$ \\
\multicolumn{1}{c}{(1)} & (2) &  (GHz) & Total &  Core & ($10^{10}$K) & (7) & Flux & Flux & (TeV) & (11) & (12) & Flux & (14) & $G_{\rm TeV}$ & (16) \\
& & (3) & (mJy) & (mJy) & (6) & & (Crab) & ($10^{-12}$ & (10) & & & ($10^{-12}$ & & (15) \\
& & & (4) & (5) & & & (8) & ph cm$^{-2}$s$^{-1}$) & & & & erg cm$^{-2}$s$^{-1}$) & & \\ 
& & & & & & & & (9) & & & & (13) & \\ [2pt] \tableline \\ [-2pt]
SHBL~J001355.9$-$185406 & 0.095       & 8.4  & 15  & 9   & 0.3   & 1   & 0.006 & 0.8         & 0.31 & 3.40 & 1    & ...   & ...  & 2.89 & ...   \\ [2pt]
KUV~00311$-$1938        & 0.506$^{a}$ & 8.4  & 31  & 26	 & 4.3	 & 1   & 0.010 & 1.2$^{a}$   & 0.33 & 3.70 & 2    & 40.0  & 1.76 & 3.17 & ...   \\ [2pt]
1ES~0033+595            & 0.240$^{a}$ & 8.4  & 59  & 44	 & 1.5	 & 1   & 0.015 & 5.5$^{b}$   & 0.15 & 3.80 & 3,4  & 28.6  & 1.87 & 2.36 & ...   \\ [2pt]
RGB~J0136+391           & 0.400$^{a}$ & 8.4  & 40  & 29	 & ur    & 1   & ...   & ...$^{c}$   & ...  & ...  & 4    & 61.9  & 1.69 & ...  & ...   \\ [2pt]
RGB~J0152+017           & 0.080	      & 8.4  & 51  & 43	 & 2.9	 & 1   & 0.020 & 2.7         & 0.30 & 2.95 & 5    & 9.6   & 1.79 & 2.88 & ...   \\ [2pt]
1ES~0229+200            & 0.140	      & 8.4  & 28  & 20	 & 3.6   & 1   & 0.017 & 2.3         & 0.30 & 2.59 & 6    & ...   & ...  & 3.19 & ...   \\ [2pt]
PKS~0301$-$243          & 0.266	      & 15.4 & 218 & 157 & 5.2   & 2   & 0.014 & 3.3         & 0.20 & 4.60 & 7    & 76.6  & 1.94 & 1.83 & ...   \\ [2pt]
IC~310                  & 0.019	      & 15.4 & 102 & 62	 & 2.0   & 2   & 0.023 & 3.1         & 0.30 & 2.00 & 8    & 9.8   & 2.10 & 2.86 & ...   \\ [2pt]
RBS~0413                & 0.190	      & 8.4  & 22  & 17	 & 4.5   & 1   & 0.009 & 1.5         & 0.25 & 3.18 & 9    & 17.0  & 1.55 & 2.88 & ...   \\ [2pt]
1ES~0347$-$121          & 0.188	      & 8.4  & 9   & 7   & 0.6   & 1   & 0.022 & 3.9         & 0.25 & 3.10 & 10   & ...   & ...  & 3.67 & ...   \\ [2pt]
1ES~0414+009            & 0.287	      & 8.4  & 49  & 36	 & 0.8	 & 1   & 0.021 & 5.2         & 0.20 & 3.40 & 11   & 7.8   & 1.98 & 2.88 & ...   \\ [2pt]
PKS~0447$-$439          & 0.200$^{b}$ & ...  & ... & ... & ...   & ... & 0.027 & 4.7         & 0.25 & 3.89 & 12   & 135.6 & 1.86 & ...  & ...   \\ [2pt]
1ES~0502+675            & 0.314$^{c}$ & 8.4  & 23  & 17	 & 0.4	 & 1   & 0.060 & 8.1$^{b}$   & 0.30 & 3.92 & 13   & 42.3  & 1.49 & 3.85 & ...   \\ [2pt]
PKS~0548$-$322          & 0.069	      & 8.4  & 27  & 20	 & 0.3   & 1   & 0.015 & 2.7         & 0.25 & 2.86 & 14   & ...   & ...  & 3.01 & ...   \\ [2pt]
RX~J0648.7+1516         & 0.179	      & 8.4  & 43  & 34	 & ur    & 1   & 0.033 & 8.1$^{a}$   & 0.20 & 4.40 & 15   & 20.4  & 1.74 & 2.85 & ...   \\ [2pt]
1ES~0647+250            & 0.450	      & 8.4  & 53  & 42	 & 3.2   & 1   & 0.030 & 19.5$^{b}$  & 0.10 & ...  & 16   & 27.2  & 1.59 & 2.91 & ...   \\ [2pt]
RGB~J0710+591           & 0.125	      & 8.4  & 40  & 27	 & 1.5	 & 1   & 0.029 & 3.9         & 0.30 & 2.69 & 17   & 13.3  & 1.53 & 3.23 & 0.064 \\ [2pt]
1ES~0806+524            & 0.138	      & 22.2 & 89  & 64	 & 0.9   & 3   & 0.016 & 2.2         & 0.30 & 3.60 & 18   & 27.9  & 1.94 & 2.53 & 0.120 \\ [2pt]
RBS~0723                & 0.198	      & 8.4  & 6   & 6   & ...   & 4   & 0.025 & 6.1$^{b}$   & 0.20 & ...  & 19   & 9.3   & 1.48 & 3.83 & ...   \\ [2pt]
1RXS~J101015.9$-$311909 & 0.143	      & 8.4  & 35  & 30	 & 0.6   & 1   & 0.010 & 2.4         & 0.20 & 3.08 & 20   & 9.8   & 2.24 & 2.64 & ...   \\ [2pt]
1ES~1011+496            & 0.212	      & 15.4 & 191 & 106 & 2.5   & 3   & 0.065 & 15.8        & 0.20 & 4.00 & 21   & 72.6  & 1.85 & 2.59 & ...   \\ [2pt]
1ES~1101$-$232          & 0.186	      & 8.4  & 28  & 23	 & 0.6   & 5   & 0.019 & 4.5         & 0.20 & 2.94 & 22   & 6.1   & 1.80 & 3.10 & ...   \\ [2pt]
Markarian~421           & 0.031	      & 8.6  & 421 & 285 & 25.6	 & 6   & 0.645 & 156.7$^{d}$ & 0.20 & 2.20 & 23   & 375.7 & 1.77 & 3.71 & 0.083 \\ [2pt]
Markarian~180           & 0.045	      & 22.2 & 83  & 40	 & 7.6	 & 5   & 0.093 & 22.5        & 0.20 & 3.30 & 24   & 14.9  & 1.74 & 3.10 & 0.094 \\ [2pt]
RX~J1136.5+6737         & 0.134	      & 8.4  & 23  & 19	 & ur	 & 7   & 0.015 & 3.6$^{b}$   & 0.20 & ...  & 25   & 8.4   & 1.68 & 2.97 & ...   \\ [2pt]
1ES~1215+303            & 0.130	      & 15.4 & 295 & 224 & ur	 & 2   & 0.032 & 7.7         & 0.20 & 2.96 & 26   & 61.2  & 2.02 & 2.26 & 0.087 \\ [2pt]
1ES~1218+304	        & 0.184	      & 8.4  & 39  & 24	 & 1.5	 & 5   & 0.050 & 12.2        & 0.20 & 3.08 & 27   & 37.8  & 1.71 & 3.34 & ...   \\ [2pt]
MS~1221.8+2452	        & 0.218	      & 8.4  & 21  & 16	 & 0.6	 & 1   & 0.040 & 9.7$^{b}$   & 0.20 & ...  & 28   & 7.0   & 2.03 & 3.50 & ...   \\ [2pt]
1ES~1312$-$423	        & 0.105	      & ...  & ... & ... & ...   & ... & 0.007 & 1.1$^{a}$   & 0.28 & 2.85 & 29   & ...   & ...  & ...  & ...   \\ [2pt]
\end{tabular}}
\end{center}
\end{sidewaystable*}

\begin{sidewaystable*}
\begin{center}
\vspace*{3.5in}
{\footnotesize Table 7 (cont.): VLBI and Gamma-Ray Properties of the TeV HBLs \\
\begin{tabular}{l c c c c c c c c c c c c c c c} \tableline \tableline \\ [-4pt]
\multicolumn{1}{c}{Source} & $z$ & $\nu$ & VLBI & VLBI & $T_{B}$ & Ref & TeV & TeV & Cutoff & Ind & Ref & {\em Fermi} & Ind & Log & $\overline{m}$ \\
\multicolumn{1}{c}{(1)} & (2) &  (GHz) & Total &  Core & ($10^{10}$K) & (7) & Flux & Flux & (TeV) & (11) & (12) & Flux & (14) & $G_{\rm TeV}$ & (16) \\
& & (3) & (mJy) & (mJy) & (6) & & (Crab) & ($10^{-12}$ & (10) & & & ($10^{-12}$ & & (15) \\
& & & (4) & (5) & & & (8) & ph cm$^{-2}$s$^{-1}$) & & & & erg cm$^{-2}$s$^{-1}$) & & \\ 
& & & & & & & & (9) & & & & (13) & \\ [2pt] \tableline \\ [-2pt]
PKS~1424+240            & 0.604$^{a}$ & 15.4 & 218 & 123 & 6.1	 & 2   & 0.042 & 21.0        & 0.12 & 3.80 & 30   & 145.0 & 1.78 & 2.41 & 0.069 \\ [2pt]
H~1426+428              & 0.129	      & 8.4  & 22  & 19	 & 1.1	 & 8   & 0.136 & 20.4        & 0.28 & 3.55 & 31   & 16.8  & 1.32 & 4.02 & ...   \\ [2pt]
1ES~1440+122            & 0.163	      & 8.4  & 22  & 17	 & ur    & 1   & 0.010 & 2.4$^{b}$   & 0.20 & 3.40 & 13   & 8.9   & 1.41 & 2.79 & ...   \\ [2pt]
PG~1553+113             & 0.500$^{d}$ & 22.2 & 134 & 95	 & 1.2	 & 5   & 0.080 & 29.3$^{b}$  & 0.15 & 4.27 & 32   & 197.4 & 1.67 & 2.87 & 0.082 \\ [2pt]
Markarian~501           & 0.034	      & 8.3  & 902 & 476 & 51.9	 & 9   & 0.229 & 31.1$^{d}$  & 0.30 & 2.72 & 33   & 114.4 & 1.74 & 2.70 & 0.037 \\ [2pt]
H~1722+119              & 0.170$^{a}$ & 8.4  & 70  & 63	 & ur    & 1   & 0.020 & 8.1$^{b}$   & 0.14 & ...  & 34   & 42.3  & 1.93 & 2.52 & 0.114 \\ [2pt]
1ES~1727+502            & 0.055	      & 15.4 & 101 & 68	 & 5.8	 & 2   & 0.021 & 7.7$^{b}$   & 0.15 & 3.20 & 35,4 & 9.7   & 1.83 & 2.31 & ...   \\ [2pt]
1ES~1741+196            & 0.083	      & 8.4  & 145 & 95	 & 2.9	 & 1   & 0.008 & 1.4$^{b}$   & 0.25 & ...  & 36   & 9.4   & 1.62 & 1.93 & ...   \\ [2pt]
HESS~J1943+213	        & 0.140$^{e}$ & 1.6  & 31  & 31	 & 0.006 & 10  & 0.018 & 1.3         & 0.47 & 3.10 & 37   & ...   & ...  & 3.21 & ...   \\ [2pt]
1ES~1959+650            & 0.047	      & 15.4 & 150 & 91	 & 2.3	 & 11  & 0.146 & 19.8        & 0.30 & 2.72 & 38   & 66.9  & 1.94 & 3.30 & 0.115 \\ [2pt]
PKS~2005$-$489          & 0.071       &	8.6  & 461 & 454 & 1.6   & 7   & 0.029 & 2.6         & 0.40 & 3.20 & 39   & 48.1  & 1.78 & 2.13 & ...   \\ [2pt] 
PKS~2155$-$304	        & 0.116	      & 15.4 & 181 & 139 & 2.2   & 11  & 0.178 & 43.2        & 0.20 & 3.53 & 40   & 282.8 & 1.84 & 3.05 & ...   \\ [2pt]
B3~2247+381             & 0.119	      & 8.4  & 55  & 40	 & 2.7   & 1   & 0.021 & 5.0         & 0.20 & 3.20 & 41   & 13.2  & 1.84 & 2.72 & ...   \\ [2pt]
1ES~2344+514            & 0.044	      & 15.4 & 118 & 87	 & 5.4	 & 11  & 0.078 & 10.6        & 0.30 & 2.78 & 42   & 20.6  & 1.72 & 3.11 & ...   \\ [2pt]
H~2356$-$309            & 0.165	      & 8.4  & 24  & 17	 & 4.4   & 5   & 0.016 & 3.1         & 0.24 & 3.06 & 43   & 7.1   & 1.89 & 3.10 & ...   \\ [2pt] \tableline
\end{tabular}}
\end{center}
{\footnotesize {\bf Notes.}
Column~1: `Canonical Name' from TeVCat, Column~2: redshift, Column~3: VLBI observing frequency, Column~4: total VLBI flux density,
Column~5: VLBI core flux density, Column~6: core observer-frame Gaussian brightness temperature (ur=unresolved), Column~7: reference for VLBI data,
Column~8: integrated TeV photon flux above the cutoff energy in multiples of the Crab flux,
Column~9: integrated TeV photon flux above the cutoff energy, Column~10: cutoff energy for TeV flux, Column~11: TeV photon spectral index,
Column~12: reference for TeV data, Columns~13 and 14: 2FGL {\em Fermi} energy flux and photon spectral index.
Column~15: log TeV loudness (see $\S$~\ref{loudness}).
Column~16: intrinsic modulation index from Richards et al.\ (2014) (see $\S$~\ref{modulation}).}\\ 
{\footnotesize {\bf Notes for Column 2.}
$a$: Value is a lower limit. $b$: Uncertain. Value used is from Prandini et al.\ 2012.
$c$: Uncertain. Value used is from NED, but is controversial. $d$: range 0.43 to 0.58. We use the mean.
$e$: Value is a lower limit, but nature of source is controversial. See discussion later in text.}\\
{\footnotesize {\bf Notes for Column 9.}
$a$: Computed from a differential flux from the reference. $b$: Computed from a flux in Crabs from the reference.
$c$: Positive detection reported, no other information. $d$: Mean value computed from multiple fluxes in the reference.}\\
{\footnotesize {\bf References for Column 7.}
(1) This paper; (2) MOJAVE program; (3) Piner \& Edwards 2013; (4) Bourda et al.\ 2010; (5) Tiet et al.\ 2012;
(6) Piner et al.\ 2012; (7) {\url http://astrogeo.org/}; (8) Piner et al.\ 2008; (9) Piner et al.\ 2007; (10) Gab{\'a}nyi et al.\ 2013;
(11) Piner \& Edwards 2004}\\ 
{\footnotesize {\bf References for Column 12.}
(1) Abramowski et al.\ 2013a; (2) Becherini et al.\ 2012; (3) Mariotti 2011; (4) Mazin 2012; (5) Aharonian et al.\ 2008;
(6) Aliu et al.\ 2014; (7) Abramowski et al.\ 2013b; (8) Aleksi{\'c} et al.\ 2010; (9) Aliu et al.\ 2012a; (10) Aharonian et al.\ 2007a;
(11) Aliu et al.\ 2012b; (12) Abramowski et al.\ 2013d; (13) Benbow 2011; (14) Aharonian et al.\ 2010; (15) Aliu et al.\ 2011;
(16) De Lotto 2012; (17) Acciari et al.\ 2010; (18) Acciari et al.\ 2009a; (19) Mirzoyan 2014a; (20) Abramowski et al.\ 2012;
(21) Albert et al.\ 2007a; (22) Aharonian et al.\ 2007b; (23) Albert et al.\ 2007b; (24) Albert et al.\ 2006a; (25) Mirzoyan 2014b;
(26) Aleksi{\'c} et al.\ 2012a; (27) Acciari et al.\ 2009b; (28) Cortina 2013a; (29) Abramowski et al.\ 2013c (30) Archambault et al.\ 2014;
(31) Horan et al.\ 2002; (32) Aleksi{\'c} et al.\ 2012b; (33) Acciari et al.\ 2011b; (34) Cortina 2013b; (35) Aleksi{\'c} et al.\ 2014b;
(36) Berger 2011; (37) Abramowski et al.\ 2011; (38) Albert et al.\ 2006b; (39) Acero et al.\ 2010; (40) Abramowski et al.\ 2010a;
(41) Aleksi{\'c} et al.\ 2012c; (42)Acciari et al.\ 2011a; (43) Abramowski et al.\ 2010b}
\end{sidewaystable*}

About half of the VLBI data in Table~\ref{alltab} (20 sources) comes from this paper; VLBI data
for most of the other HBLs comes from either our prior publications (13 sources),
or from the MOJAVE program (5 sources). Four sources have VLBI data taken from elsewhere in the literature.
Only two of the 44 sources (both below $-40\arcdeg$ declination) have 
no VLBI data in the literature.
All of the VLBI data taken from elsewhere, such as that taken from the MOJAVE survey, has been independently model-fit
by us if the visibility data files were available online; if not then published values have been used.
For sources with multiple epochs of VLBI data, we have model-fit all epochs, and then used the epoch having
the median core brightness temperature. VLBI data at an observing frequency of 8~GHz
(the observing frequency used for this paper) was preferred if it was available; if not, then
data at either 15 or 22 ~GHz (or in a single case, 1.6~GHz) has been used.
References for all VLBI data used are given in the notes to Table~\ref{alltab}.

The {\em Fermi} gamma-ray fluxes and spectral indices in Table~\ref{alltab} are taken from the 2FGL catalog (Nolan et al.\ 2012).
Only six of the 44 sources have not been detected by {\em Fermi}, as of the 2FGL catalog.
For the TeV gamma-ray data, an integrated photon flux, cutoff energy for that flux, 
and spectral index (uncorrected for Extra-galactic Background Light [EBL] absorption) was taken from the literature, if available.
That integrated photon flux was then independently converted to a multiple of the Crab nebula flux
using the Crab spectrum from Aharonian et al.\ (2006); this may cause slight differences from Crab fluxes
quoted in the original papers referenced in Table~\ref{alltab}.
In some cases, only a flux that was already expressed in multiples of the Crab flux was given in the literature.
In those cases, the integrated photon flux above the cutoff energy
was calculated from that using the Crab spectrum from Aharonian et al.\ (2006).
For many of the sources, the numbers given in Table~\ref{alltab} match the flux in multiples of the Crab flux, cutoff energy, and spectral index
quoted for that source in TeVCat; but in a number of cases they differ, due either to different literature sources used, or
differing Crab nebula standards. 
Many of the newly discovered sources have only a single flux value in the literature, but
for frequently observed sources, the variability of the TeV HBLs makes selection of 
a single flux value problematic. Because sources are observed over different time and energy ranges with different instruments,
calculation of a formal mean would be difficult. Nevertheless, we have tried to select
a `typical' flux value for variable sources, excluding extreme high or low states. 
This exclusion of extreme high or low states
may also cause numbers in Table~\ref{alltab} to differ from those in TeVCat. 
In any event, the TeV data will almost certainly not be contemporaneous with the VLBI measurements.
References for all TeV data used are given in the notes to Table~\ref{alltab}.

\subsection{VLBI Flux Densities and Brightness Temperatures}
\label{sandtb}
A histogram of the VLBI core flux densities of the TeV HBLs from Table~\ref{alltab}, 
which is indicative of the most compact emission from these sources, is shown in Figure~3.
New sources with VLBI data from this paper are shown in yellow, while
sources with data taken from elsewhere are shown in blue.
The range in core flux densities spans from a few mJy (e.g., 1ES~0347$-$121) to
a few hundred mJy (e.g., Mrk~421 and Mrk~501), with a median of 38 mJy.
Note that all of the new sources added in this paper have cores that are under 100 mJy. 
As the TeV gamma-ray telescopes have become more sensitive and begun to
detect fainter objects, these sources have also tended to be fainter in the radio,
a potential correlation that is explored in $\S$~\ref{correlations}.

\begin{figure*}[!t]
\centering
\includegraphics[angle=90,scale=0.45]{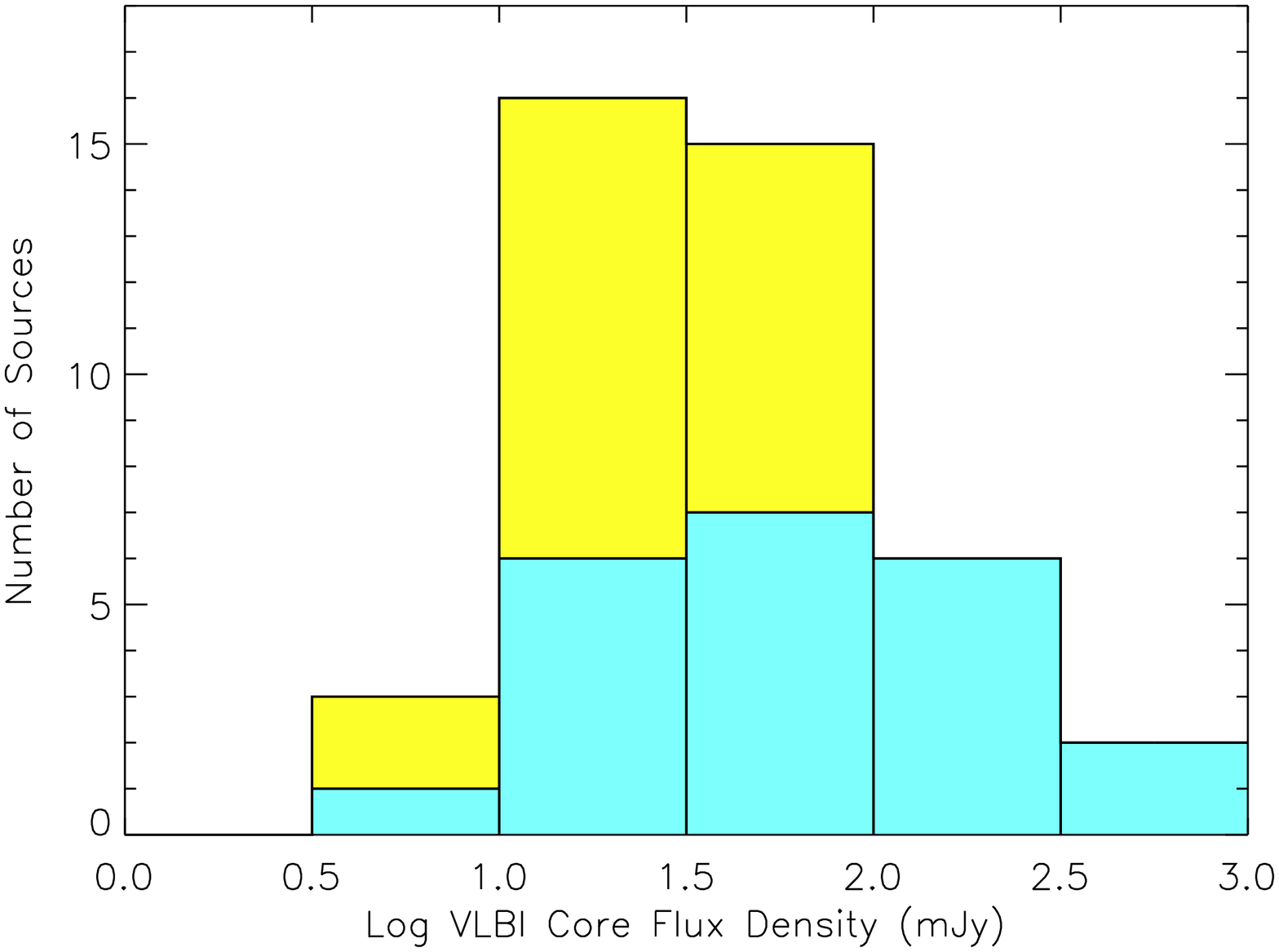}
\caption{Histogram of VLBI core flux densities of TeV HBLs from Table~\ref{alltab}.
New sources with VLBI data from this paper are shown in yellow (20 sources).
Sources with data taken from elsewhere are shown in blue (22 sources).}
\end{figure*}

\begin{figure*}[!ht]
\centering
\includegraphics[angle=90,scale=0.45]{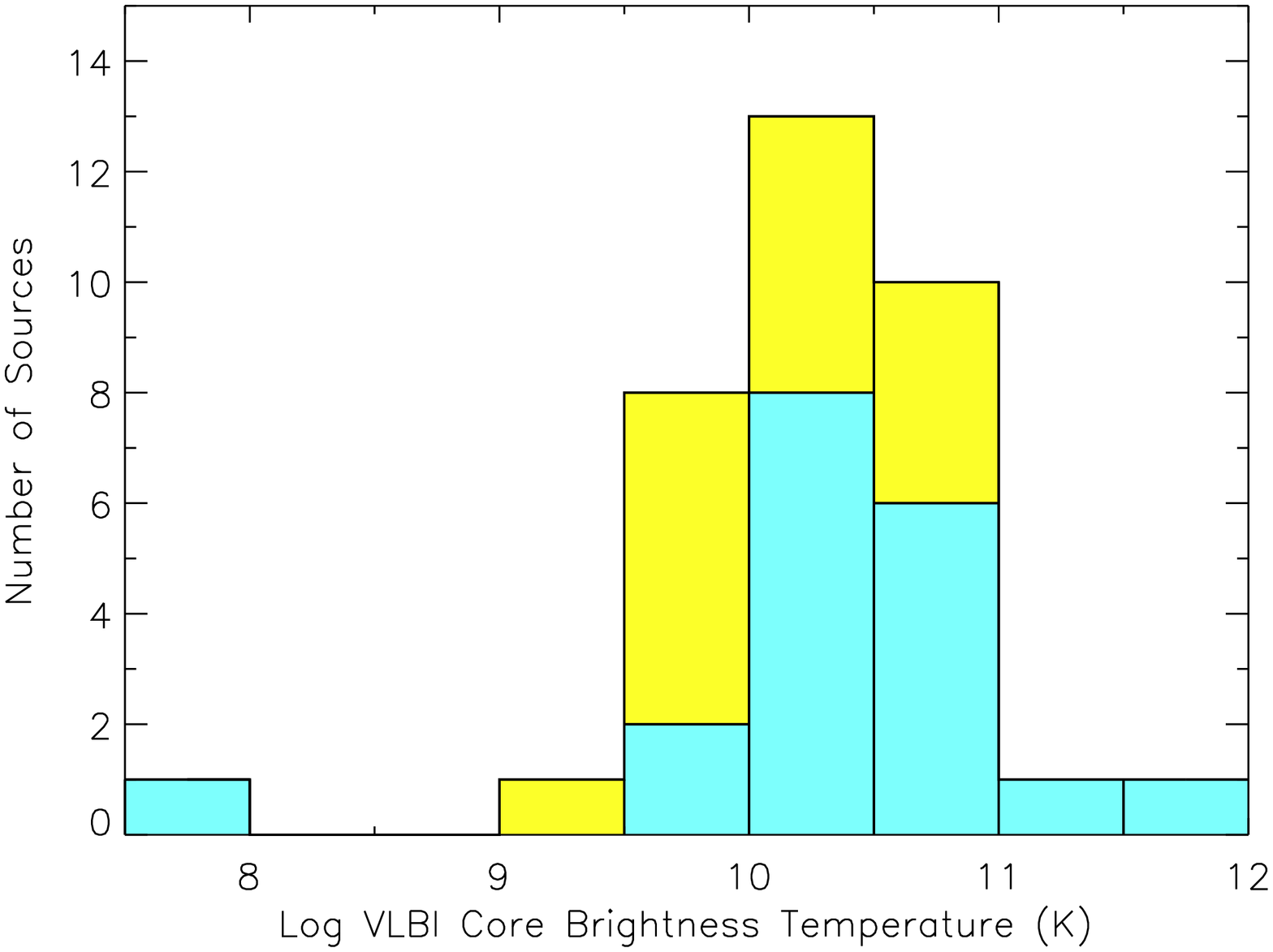}
\caption{Histogram of observer-frame Gaussian core brightness temperatures of TeV HBLs from Table~\ref{alltab},
for sources whose best-fit core size is not zero.
New sources with VLBI data from this paper are shown in yellow (16 sources).
Sources with data taken from elsewhere are shown in blue (19 sources).
The outlier is HESS~J1943+213 (see text).}
\end{figure*}

A histogram of the VLBI core brightness temperatures from Table~\ref{alltab} is shown in Figure~4.
New sources with VLBI data from this paper are shown in yellow, while
sources with data taken from elsewhere are shown in blue.
A brightness temperature value has been plotted in Figure~4 unless the best-fit value for the 
core size is zero (indicated by `ur' in Table~\ref{alltab}).
However, as indicated by the brightness temperature error analysis done for the 20 sources
observed for this paper in $\S$~\ref{tb}, some of these brightness temperature values
are probably actually lower limits. The median brightness temperature in Figure~4
is $2\times10^{10}$~K, which is the same as the typical brightness temperature obtained
from the brightness temperature error analysis in $\S$~\ref{tb}.
See $\S$~\ref{tb} for discussion of the physical interpretation of such brightness temperatures
in terms of intrinsic brightness temperature limits and relativistic beaming.

Some outliers are notable in Figure~4. The only two TeV HBLs with brightness temperatures over
$10^{11}$~K are the well-studied sources Mrk~421 and Mrk~501.
The low brightness temperature outlier, with a measured brightness temperature of only $6\times10^{7}$~K,
is the source HESS~J1943+213 which lies close to the Galactic plane.
This source was observed with the European VLBI Network (EVN) at 1.6~GHz by Gab{\'a}nyi et al.\ (2013),
who measured it to have a flux density of 31 mJy and an angular size of 16 mas, 
giving it a brightness temperature two orders of magnitude lower than all other TeV HBLs in Figure~4.
The distribution in Figure~4 casts significant doubt on the HBL classification of this object
(unless it is affected by an unusually large amount of interstellar scattering, see the discussion in Gab{\'a}nyi et al.\ 2013),
and Gab{\'a}nyi et al.\ (2013) suggest instead a galactic origin for this source, in the form of a remote pulsar wind nebula (PWN).
This interpretation may be strengthened by the lack of detection of any significant variability from this
object from radio to TeV gamma-rays (Abramowski et al.\ 2011).

\subsection{TeV Loudness}
\label{loudness}
In this section, we quantify the distribution of the ratio of TeV gamma-ray to radio
luminosity present in the TeV HBL population. Lister et al.\ (2011) performed a similar
analysis for {\em Fermi}-detected blazars by defining a quantity that they called
the gamma-ray loudness, $G_{r}$. This quantity was defined by Lister et al.\ (2011) as the ratio of the gamma-ray
luminosity between 0.1~GeV and 100~GeV, divided by the radio luminosity over a 15~GHz wide
bandwidth, calculated from the VLBA flux density at 15~GHz (see equations 2 through 4 of Lister et al.\ 2011).
We make straightforward modifications to Equations (2) though (4) of Lister et al.\ (2011) to
adapt their gamma-ray loudness statistic to the TeV energy range 
and main VLBA observing frequency considered in this paper.
We calculate the TeV loudness, $G_{\rm TeV}=L_{\rm TeV}/L_{\rm R}$, using the gamma-ray luminosity between 0.3 and 30~TeV,
accounting for the different lower energy thresholds for the different measurements given in Table~\ref{alltab}.
The modified versions of Equations (2) though (4) from Lister et al.\ (2011) are:

\begin{equation}
S_{\rm TeV}=\frac{(\Gamma-1)C_{1}E_{0}F_{0}}{(\Gamma-2)}\Bigg(\frac{E_{0}}{E_{1}}\Bigg)^{\Gamma-2}\Bigg[1-\Bigg(\frac{E_{1}}{E_{2}}\Bigg)^{\Gamma-2}\Bigg],
\end{equation}
where $F_{0}$ is the measured photon flux above the cutoff energy $E_{0}$, $\Gamma$ is the photon spectral index,
$E_{1}=0.3$~TeV, $E_{2}=30$~TeV, and $C_{1}=1.602$~erg~TeV$^{-1}$, and $S_{\rm TeV}$ is in erg~cm$^{-2}$~s$^{-1}$,

\begin{equation}
L_{\rm TeV}=\frac{4\pi D_{L}^{2}S_{\rm TeV}}{(1+z)^{2-\Gamma}}~\rm{erg~s^{-1},}
\end{equation}
where $D_{L}$ is the luminosity distance in cm, and

\begin{equation}
L_{\rm R}=\frac{4\pi D_{L}^{2}\nu S_{\nu}}{(1+z)}~\rm{erg~s^{-1},}
\end{equation}
where $S_{\nu}$ is the total VLBA flux density in erg~cm$^{-2}$~s$^{-1}$~GHz$^{-1}$, and $\nu=8$~GHz.
The quantities $F_{0}$, $E_{0}$, $\Gamma$, $S_{\nu}$, and $z$ are tabulated in Table~\ref{alltab}.
If a photon spectral index was not measured for a source, then we used the median measured
photon spectral index of $\Gamma=3.2$. We assume a flat radio spectral index ($\alpha=0$)
for the radio $k$-correction and luminosity calculation.

The TeV loudness is tabulated in Table~\ref{alltab}, and a histogram of this statistic is shown in Figure~5.
As can be seen from Figure~5, the distribution
spans about two orders of magnitude, from about 10$^{2}$ to 10$^{4}$,
but the distribution is peaked around the median value of about 10$^{3}$.
A similar range of about two orders of magnitude in gamma-ray loudness
is spanned by the BL~Lac objects studied by Lister et al.\ (2011), 
although much of the range in gamma-ray loudness observed by those authors is due to a
mix of HBLs, IBLs, and LBLs in the MOJAVE sample.
The fact that a similar range is observed here among just the TeV HBLs is mostly due to
the inclusion of the radio-faintest TeV HBLs at flux-density levels of a few millijanskys.
For example, the single source with TeV loudness greater than 10$^{4}$ in Figure~5 is the extreme blazar H~1426+428, 
which is among the brighter TeV sources, but has a VLBA flux density of only about 20~mJy.
Conversely, the two sources with TeV loudness less than 10$^{2}$ in Figure~5 are the relatively radio-bright HBLs
PKS~0301$-$243 and 1ES~1741+196.

\begin{figure*}[!t]
\centering
\includegraphics[angle=90,scale=0.45]{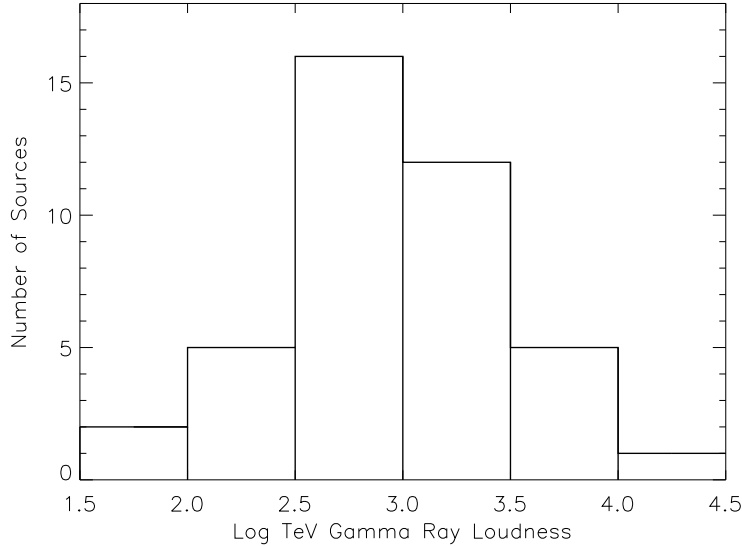}
\caption{Histogram of the TeV loudness for the 41 TeV HBLs from Table~\ref{alltab} 
with both TeV and VLBI fluxes. The TeV loudness is defined in $\S$~\ref{loudness}.}
\end{figure*}

We might expect there to be a significant anti-correlation between TeV loudness and redshift,
because of EBL absorption of TeV gamma-rays from distant sources.
However, a correlation analysis of these two quantities does not yield a significant correlation,
possibly because the vast majority of the TeV HBLs are clustered at low redshifts,
and these low redshift sources already show a large intrinsic scatter in TeV loudness.
Lister et al.\ (2011) found a significant anti-correlation between gamma-ray loudness
and {\em Fermi} photon spectral index for the BL~Lac objects in their sample.
We confirm this correlation for the TeV HBLs in this paper at a marginally significant level;
for the 35 sources in Table~\ref{alltab} with both measured TeV loudness and photon index 
a partial Spearman rank correlation test (excluding effects of redshift) has a significance of 0.03.
See Lister et al.\ (2011) for a discussion of the implications of such a correlation for
emission models in BL~Lac objects.

\subsection{Flux-Flux Correlations}
\label{correlations}
Establishing whether or not blazar fluxes in different wavebands (e.g., radio and gamma-ray) are intrinsically
correlated, independent of any common-distance effects, is important to establishing 
to what degree the emission regions in the jet at these different wavebands are connected.
The existence of correlations implies that the emission regions,
even if they occur in different components of the jet with different
beaming parameters, are related through some physical property of the source. 
Truly uncorrelated fluxes would instead imply that the emission regions probed by radio and gamma-ray observations are 
completely independent of each other.

\begin{figure*}[!t]
\centering
\includegraphics[scale=0.65]{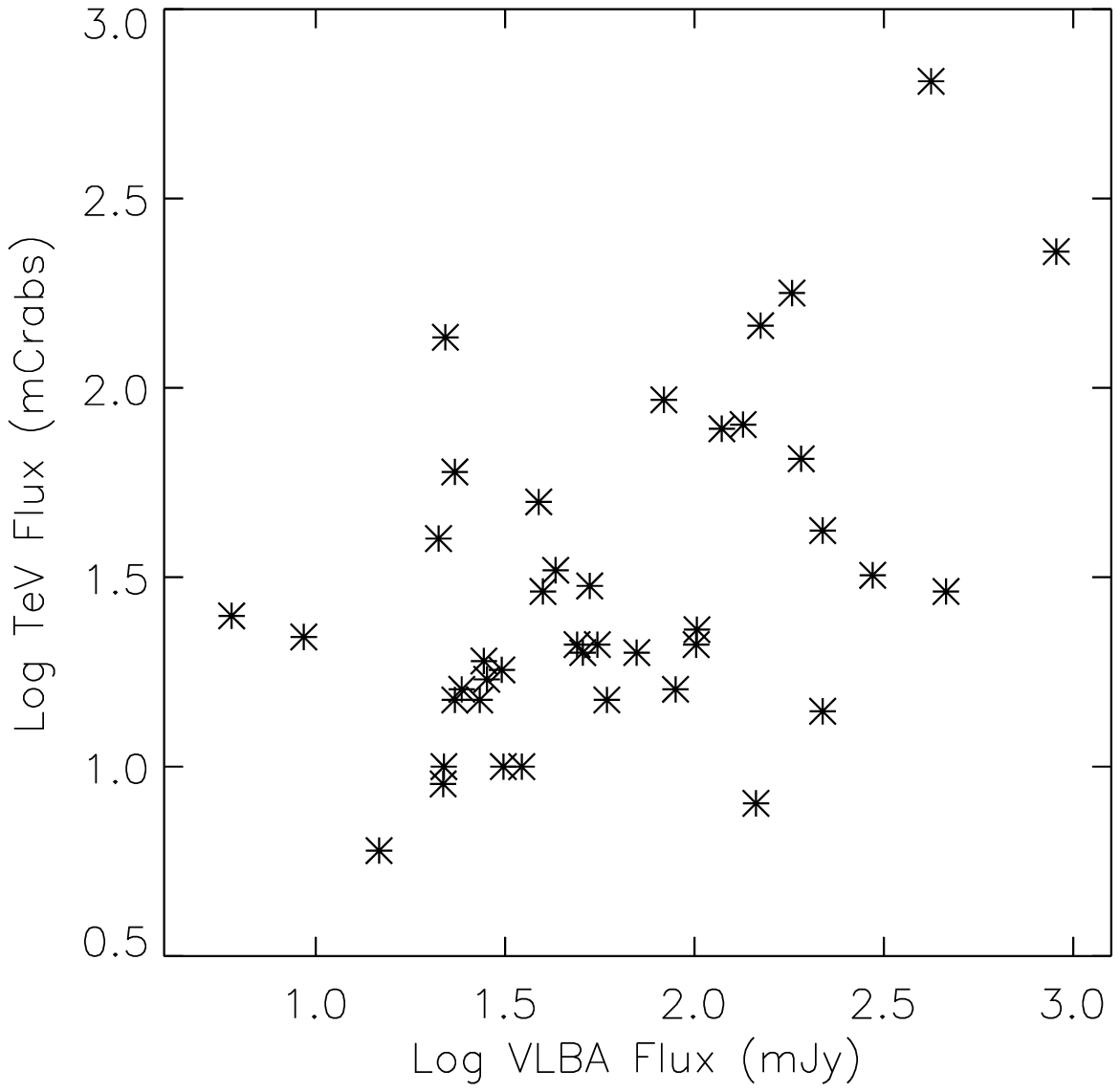}
\includegraphics[scale=0.65]{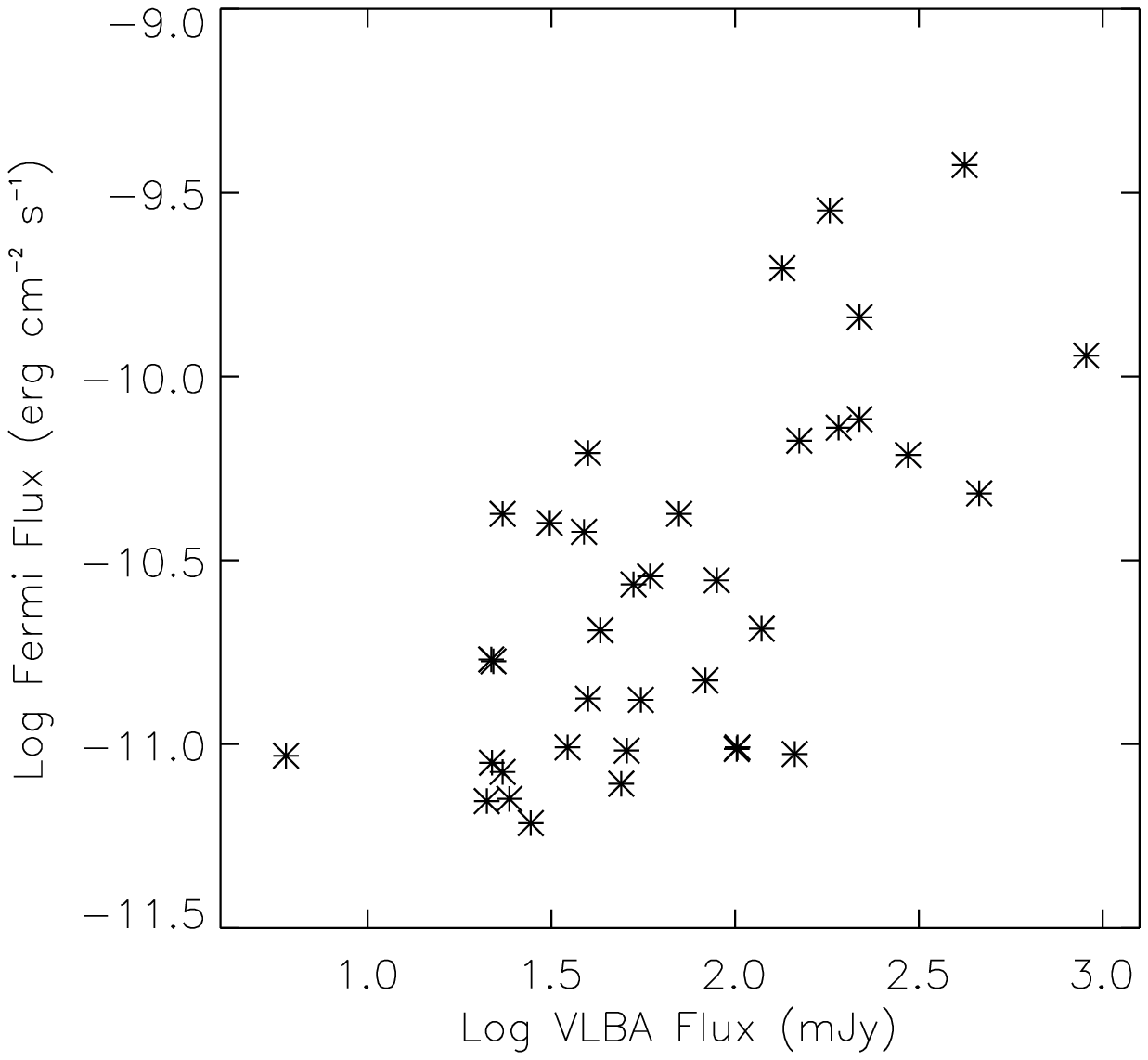}
\caption{Top panel: Plot of TeV flux in milliCrabs vs. total VLBA flux density in millijanskys
(41 sources). Bottom panel: Plot of 2FGL {\em Fermi} flux vs. total VLBA flux density in millijanskys
(37 sources). All flux values are from Table~\ref{alltab}.}
\end{figure*}

The top panel of Figure~6 shows the TeV flux in milliCrabs versus the total VLBI flux density
in millijanskys for the 41 sources in Table~\ref{alltab} with measured TeV and VLBI fluxes.
A partial correlation analysis (excluding effects of redshift)
gives a Pearson partial correlation coefficient of 0.50, with a significance of $9\times10^{-4}$ (99.91\% chance of correlation).
Repeating the analysis using the VLBI model-fit core flux density instead of the total flux density
yields a similar but slightly lower Pearson partial correlation coefficient of 0.48, with a significance
of $1.7\times10^{-3}$ (99.83\% chance of correlation). The correlation with core flux density is probably slightly
less significant because the extra step of model fitting introduces some scatter into the core flux density values,
particularly when there is a bright jet component close to the core.
The high value for the significance of the correlation shown in the top panel of Figure~6
is partly due to the two sources Mrk~421 and Mrk~501, which
are bright in both the radio and TeV gamma-rays.
However, even if those two sources are excluded (and note that there is no particular physical reason for excluding them),
the partial correlation remains significant, although at a lower level of 96.3\%.
This is the first time, to our knowledge, that a correlation between the TeV flux and the VLBI flux has been
established for the TeV HBL population.

To further test the robustness of this new correlation, we employed Monte Carlo simulations
using the method described by Pavlidou et al.\ (2012) to generate intrinsically uncorrelated
permutations of the data for comparison purposes, using the monochromatic flux density at 0.3~TeV 
computed from Table~\ref{alltab} as the TeV flux sample.
Because this method requires applying {\em k}-corrections to the permuted data, the six
sources without a TeV photon spectral index in Table~\ref{alltab} were excluded.
A comparison of $10^{7}$ randomly permuted datasets for the 35 remaining sources with the actual dataset yields a significance
of correlation of 0.056 (94.4\% chance of correlation). This result is now only marginally significant; however, Pavlidou et al.\ (2012)
state that their method is conservative for small samples such as this, and that existing intrinsic correlations
may not be verified. We also note that the non-contemporaneous nature of the data may
wash out a stronger correlation that might have existed in concurrently measured data.

The bottom panel of Figure~6 shows the 2FGL {\em Fermi} flux versus the total VLBI flux density
for the 37 sources in Table~\ref{alltab} with measured {\em Fermi} and VLBI fluxes.
A similar plot is shown for the MOJAVE program sources in Figure~1 of Lister et al.\ (2011), and the bottom panel
of Figure~6 basically continues the trend for the HBLs shown in that figure toward lower VLBI and {\em Fermi} flux values.
Correlations between radio and {\em Fermi} fluxes for larger samples of {\em Fermi}
blazars have been established by a number of authors (e.g., Kovalev et al.\ 2009;
Ackermann et al.\ 2011; Linford et al.\ 2012). A correlation between {\em Fermi} and radio fluxes
solely for the TeV HBL sub-population of {\em Fermi} sources was claimed by Xiong et al.\ (2013),
although they did not address common-distance effects.
We confirm such a correlation between the {\em Fermi} and VLBI fluxes of the TeV HBLs; 
a partial correlation analysis (excluding effects of redshift)
gives a Pearson partial correlation coefficient of 0.76, with a formal significance of about $10^{-7}$. 
However, we note that because this sample is TeV-selected rather than {\em Fermi}-selected, 
such tests may overestimate the significance of correlations because they do not address upper
limits for the TeV HBLs that are not in the 2FGL catalog, although this is a small number of objects (6 sources).

\subsection{Radio Variability and Modulation Indices}
\label{modulation}
The variability of a radio source is an important property
(potentially constraining both relativistic beaming
and the relative locations of emission regions at different wavebands) that cannot be well-studied
by sequences of a only a few VLBA images.
The largest current effort to study radio variability of blazars is the
Owens Valley Radio Observatory (OVRO) monitoring program\footnote{\url{http://www.astro.caltech.edu/ovroblazars}}, which
presently monitors more than 1800 blazars about twice per week.

The OVRO program's chosen parameter to characterize variability is the
`intrinsic modulation index', $\overline{m}$, which is an estimate of the standard deviation of the source
flux density divided by its mean (Richards et al.\ 2011, 2014).
Although many of the TeV HBLs are in the OVRO program, because of their relative 
faintness they are clustered near the program's measurement limits in flux density and modulation index.
For example, although 25 of the TeV HBLs in Table~\ref{alltab} are in Table~1 of Richards et al.\ (2014),
6 do not have a measured modulation index, and another 9 have a flux density and modulation index
that are excluded from the analysis for being too close to the measurement limits.
The 10 remaining high-confidence modulation indices are tabulated with the other source data in Table~\ref{alltab}.
Because only 10 TeV HBLs pass the current data cuts in the OVRO analysis, we do not attempt
correlation studies with the modulation index here, but leave it as an interesting possibility for future
study if thresholds for the OVRO variability analysis are lowered.

\section{Discussion and Conclusions}
\label{discussion}
We have investigated the parsec-scale jet structure of twenty relatively
newly discovered TeV HBLs that had not been previously well-studied with VLBI.
These newly discovered TeV HBLs extend down to only a few millijanskys
in flux density, so they are not present in large VLBI monitoring programs.
All sources were detected and imaged, and all showed parsec-scale jets that could be modeled 
with at least one Gaussian component ($\S$~\ref{mfits}). Most sources had a one-sided
core-jet morphology, although we find two cases of apparently two-sided structure ($\S$~\ref{images}).
Many sources show a common morphology of a collimated jet a few milliarcseconds long that transitions
to a lower surface brightness, more diffuse jet with a broader opening angle at a few mas from the core.
These results show that the entire TeV HBL sample, although relatively
faint in the radio, is accessible to analysis with current VLBI instruments.

As well as can be determined from only single-epoch images,
the analyses presented here support previous conclusions that Lorentz factors in
the parsec-scale cores and jets of TeV HBLs are only modestly relativistic.
We determined allowed brightness temperature ranges for each core component ($\S$~\ref{tb}),
and found that roughly half of the VLBI cores are resolved with brightness temperature
upper limits of a few times $10^{10}$~K (Table~\ref{tbtab}). A Gaussian brightness temperature of $2\times10^{10}$~K
was consistent with the data for all but one of the sources.
Such brightness temperatures do not require any relativistic beaming to
reduce them below likely intrinsic limits.
The lack of detection of counter-jets does place at least a modest limit on
the bulk Lorentz factor, although the strongest such constraint we could place was $\gamma\gtrsim2$.
The distribution of apparent opening angles ($\S$~\ref{opening}) is indistinguishable from that of the 
general gamma-ray blazar population (Pushkarev et al.\ 2009; Lister et al.\ 2011), 
so there is no indication from their jet morphology
that these sources are unusually close to the line-of-sight compared to other 
gamma-ray blazars. There is thus no evidence from these images that the slow apparent speeds of TeV 
HBLs are caused by a much closer alignment to the line-of-sight compared to the 
apparently faster sources.

The `Doppler Crisis' for TeV HBLs suggests that their parsec-scale jets are structurally more complex than those of the more
powerful blazars, and that they require at least two zones of significantly different
Lorentz factor to successfully describe them.
A consistent picture is emerging of this dichotomy in the jetted AGN population
based on multiwavelength studies of large populations, theoretical modeling,
and high-resolution imaging with VLBI.
In this picture, jets formed in a low-efficiency accretion mode typical of HBLs
(Ghisellini et al.\ 2005, 2009; Meyer et al.\ 2013b) favor interaction of
the jet walls with the external medium, causing the formation of a slow layer.
Radiative interaction between the spine and the layer may then decelerate the spine (Ghisellini et al.\ 2005),
producing longitudinal as well as transverse velocity structure, such as postulated by Georganopoulos \& Kazanas (2003).
Application of such two-zone models can also be successful in
reducing the most extreme Doppler factors sometimes required by one-zone models.
For example, for the TeV blazar PKS~1424+240, fitting the SED with a one-zone
model yields a Doppler factor of $\sim100$, while the MOJAVE VLBA data imply a 
Doppler factor of only $\sim10$ (Aleksi{\'c}, et al.\ 2014a). When the same SED is
fit by the two-zone model of Tavecchio et al.\ (2011),
the Doppler factor of the fast zone is reduced to $\sim30$, 
while that of the slower zone has the VLBA-derived value of $\sim10$ (Aleksi{\'c}, et al.\ 2014a).

VLBI imaging might detect such two-zone spine-layer jets through observations of
transverse emission structures such as limb brightening.
We do observe limb brightening in two sources (see $\S$~\ref{transverse}), although
for the majority of sources the transverse structure is either unresolved, 
or it is patchy and complex with multiple emission peaks. 
However, we note that, because of increased
source distance and lower observing frequency, the observations in this paper have
about an order of magnitude worse linear resolution compared to the 
high-frequency observations of limb-brightening in the nearby bright TeV
blazars Mrk~421 and Mrk~501 (e.g., Piner et al.\ 2009, 2010).

In the spine-layer model for TeV HBLs, the TeV emission comes from the spine while the radio emission comes from the layer,
causing different Lorentz factors to be measured in the two spectral bands. However, for
TeV radio galaxies, both the gamma-ray and the radio emission can be dominated by the layer
(Ghisellini et al.\ 2005), so that consistent Lorentz factors might be expected (no `Doppler Crisis').
VLBI observations of the three known TeV radio galaxies M87 (e.g., Hada et al.\ 2014),
Cen~A (e.g., M{\"u}ller et al.\ 2014), and 3C~84 (e.g., Nagai et al.\ 2014)
can therefore provide consistency checks on spine-layer models (e.g., Tavecchio \& Ghisellini 2008, 2014).

We extended our consideration to the full sample of TeV HBLs in $\S$~\ref{resultsall},
by combining our VLBI data on 20 sources from this paper with other VLBI and gamma-ray data from the literature.
Following the approach of Lister et al.\ (2011), we constructed a gamma-ray loudness parameter 
for the TeV HBLs ($\S$~\ref{loudness}), and found
that it spans about two orders of magnitude from extreme gamma-ray loud sources like
H~1426+428 to more radio-loud sources like PKS~0301$-$243.
There is a significant apparent partial correlation (excluding effects of redshift) between the VLBI and TeV fluxes
($\S$~\ref{correlations}), although Monte Carlo simulations using the method of Pavlidou et al.\ (2012)
showed that this correlation may intrinsically be only marginally significant. Such a correlation might
suggest that Doppler factors in different jet emission regions are correlated, even if they are significantly different,
such as occurs in the model by Lyutikov \& Lister (2010).
Note that VLBI core flares correlated with gamma-ray flares in Mrk~421 (Richards et al.\ 2013) also suggest 
a link between the VLBI emission and the gamma-ray emission in that source.

Much more information about the jet kinematics of the TeV HBLs should be revealed through the
multi-epoch VLBA monitoring of these 20 sources that is currently underway.
At least three additional epochs for each of these sources has been approved 
on the VLBA, and should be obtained over the next one to two years, in addition to high-frequency
imaging of some of the brighter TeV HBLs to investigate transverse jet structures.
When added to the 11 TeV HBLs that we have already monitored, and the 7 additional TeV HBLs
being monitored by MOJAVE, this will make 
information on parsec-scale structural changes available for $\sim90\%$ of the currently known TeV HBL population.
This is a crucial step toward understanding the jet structure of this group of sources,
as the high-energy community looks forward to many more such objects being detected by 
future TeV telescopes like the Cherenkov Telescope Array (CTA).

\vspace*{-0.10in}
\acknowledgments
The National Radio Astronomy Observatory is a facility of the National
Science Foundation operated under cooperative agreement by Associated Universities, Inc.
This research has made use of the NASA/IPAC Extragalactic Database (NED) 
which is operated by the Jet Propulsion Laboratory, California Institute of Technology, 
under contract with the National Aeronautics and Space Administration.
This research has made use of data from the MOJAVE database that is maintained by the MOJAVE team (Lister et al.\, 2009).
Part of this research was carried out at the Jet Propulsion Laboratory,
Caltech, under a contract with the National Aeronautics and Space Administration.
This work was supported by Fermi Guest Investigator grant NNX13AO82G.
We also acknowledge helpful comments from the anonymous referee.

\vspace*{0.1in}
{\it Facilities:} \facility{VLBA ()}

\end{document}